\documentclass[apj]{emulateapj}

\usepackage{natbib}
\usepackage{graphicx}
\usepackage{apjfonts}
\usepackage{epstopdf}

\newcommand{\Mpc}{$h^{-1}$\thinspace Mpc}

\newcommand{\be}{\begin{equation}}
\newcommand{\ee}{\end{equation}}

\begin{document}  
 
\shorttitle{The Sloan Great Wall}
\shortauthors{Einasto et al.}
\title{The Sloan Great Wall. Morphology and galaxy content} 
\author{M. Einasto\altaffilmark{1}, 
L. J. Liivam\"agi\altaffilmark{1}, 
E. Tempel\altaffilmark{1}, 
E. Saar\altaffilmark{1}, 
E. Tago\altaffilmark{1}, 
P. Einasto\altaffilmark{1}, 
I. Enkvist\altaffilmark{1}, 
J. Einasto\altaffilmark{1}, 
V.J. Mart\'{\i}nez\altaffilmark{2},
P. Hein\"am\"aki\altaffilmark{3},
P. Nurmi\altaffilmark{3}}

\altaffiltext{1}{Tartu Observatory, EE-61602 T\~oravere, Estonia}
\altaffiltext{2}{Observatori Astron\`omic, Universitat de Val\`encia, Apartat
de Correus 22085, E-46071 Val\`encia, Spain} 
\altaffiltext{3}{University of Turku,
Tuorla Observatory, V\"ais\"al\"antie 20, Piikki\"o, Finland} 

\begin{abstract} 

We present the results of the study of
 the morphology and galaxy content of the Sloan Great Wall (SGW), the 
richest galaxy system in the nearby Universe. We use the luminosity density field to 
determine superclusters in the SGW, and the 4th Minkowski functional $V_3$ and 
the morphological signature (the $K_1$-$K_2$ shapefinders curve) to show the 
different morphologies of the SGW, from a single filament to a multibranching, 
clumpy planar system. 
We show that the richest supercluster in the SGW, SCl~126 
and especially its core resemble a very 
rich filament, while another rich supercluster in the SGW, SCl~111, resembles a 
``multispider''---an assembly of high density regions connected by chains of 
galaxies. 
We study the substructure of individual galaxy populations determined by their color in 
these superclusters using Minkowski functionals and find that
in the high density core of the 
SGW the clumpiness of red and blue galaxies is similar, but in the outskirts of 
superclusters the distribution of red galaxies is more clumpy than that of blue 
galaxies. At 
intermediate densities, the systems of blue galaxies have 
tunnels through them. 
We assess the statistical significance of our results 
using the halo model and  smoothed bootstrap.

We study the galaxy content and the properties of groups of galaxies in  
two richest superclusters of the SGW, paying special attention to bright red 
galaxies (BRGs) and to the first ranked galaxies in SGW groups. The BRGs are the 
nearby LRGs, they are mostly bright and red  and typically reside in 
groups (several groups host 5 or more BRGs). 
About 1/3 of BRGs are spirals. The 
scatter of colors of elliptical BRGs is smaller than that of spiral BRGs. About 
half of  BRGs and of first ranked galaxies in groups have large peculiar 
velocities.
Groups with elliptical BRGs as their first ranked galaxies populate 
superclusters more uniformly than the groups, which have a spiral BRG as its 
first ranked galaxy.

The galaxy and group content of the 
core of the supercluster SCl~126  shows several differences in comparison with 
the outskirts of this supercluster and with the supercluster SCl~111. Here 
groups with BRGs are richer and have larger velocity dispersions than groups 
with BRGs in the outskirts of this supercluster and in SCl~111. The fraction of 
those BRGs which do not belong to any group is the smallest here. In the core of 
the supercluster SCl~126 the fraction of red galaxies is larger than in the 
outskirts of this supercluster or in the supercluster SCl~111.  
Here the peculiar velocities of of the first ranked galaxies are larger
than in the outskirts of this supercluster or in the supercluster SCl~111 and
the peculiar velocities of
elliptical BRGs are larger than those of spiral BRGs, while in the outskirts of 
this supercluster and in the supercluster SCl~111 the peculiar velocities of 
spiral BRGs are larger than those of elliptical BRGs. 
 
All that suggests that the formation history and evolution of individual  neighbour
superclusters in the SGW has been different, and the SGW is not a genuine physical
structure but an assembly of very rich galaxy systems. 

\end{abstract}


\section{Introduction}
\label{sec:intro}

Observations and simulations of the large scale structure of the Universe
have revealed the presence of a supercluster-void network---a network of 
galaxies, groups and clusters of galaxies connected by filaments
\citep{1978MNRAS.185..357J,1978ApJ...222..784G,zes82,
1984MNRAS.206..529E,1986ApJ...302L...1D}.
The formation of a web of galaxies and systems of 
galaxies is predicted in any physically motivated theories of the 
formation of the structure in the Universe \citep{1996Natur.380..603B}.
\citet{2010MNRAS.409..156B} emphasize that if we want to understand 
the patterns 
of the large scale structure (cosmic web)
we need to know how to quantify them. One approach for that is 
the study of the morphology of systems of galaxies.
 
Galaxies and galaxy systems form due to initial density perturbations of 
different scale. Perturbations of a scale of about 100~\Mpc\footnote{$h$ is the 
Hubble constant in units of 100~km~s$^{-1}$~Mpc$^{-1}$.} give rise to the 
largest superclusters. The largest and richest superclusters which may contain 
several tens of rich (Abell) clusters are the largest coherent systems in the 
Universe with characteristic dimensions of up to 100~\Mpc. At large scales 
dynamical evolution takes place at a slower rate and the richest superclusters 
have retained the memory of the initial conditions of their formation, and of the 
early evolution of structure \citep{1987Natur.326...48K}.

Rich superclusters had to form earlier than smaller superclusters; they are the sites 
of early star and galaxy formation \citep[e.g.][]{2005ApJ...635..832M}, 
and the first places 
where systems of galaxies form 
\citep[e.g.][]{2004A&A...424L..17V,2005ApJ...620L...1O}. Observations already 
have revealed superclusters at high redshifts 
\citep{2005MNRAS.357.1357N,2007MNRAS.379.1343S,2008ApJ...684..933G,tanaka09}.

Early supercluster catalogues have been compiled using the data on 
clusters of galaxies 
\citep{1993ApJ...407..470Z,1994MNRAS.269..301E,1995A&AS..113..451K,
e2001}.
The advent of the deep surveys of galaxies as 
\citep[2dFGRS and SDSS, see][]{col03,abaz08} started a new era 
in the studies of the large scale structure of the Universe, where systems
of galaxies can be studied in a detail impossible before.
Using these data, a number of supercluster catalogues have been compiled
\citep[][and references therein]{bas03,2003A&A...410..425E,2003A&A...405..425E,
2004MNRAS.352..939E}.

Superclusters are important traces of the  baryonic matter in
the Universe \citep{2005MNRAS.363...79G,2009MNRAS.396...53P}.
Observations have revealed the presence of warm-hot diffuse gas (WHIM) in the 
Sculptor supercluster and nearby Sculptor Wall associated with the inter-cluster 
galaxy distribution in these superclusters 
\citep[][and references therein]{2005A&A...434..801Z,2009ApJ...695.1351B}, often 
referred to as the ``missing baryons''.  Detailed structural 
and spatial information on nearby superclusters could help the targeted search 
for WHIM.
 
Gravitational lensing analysis of  supercluster
structure can be used to reconstruct the distribution of dark matter in
superclusters as has been done for the supercluster A901/902
\citep{2008MNRAS.385.1431H}.

Rich superclusters contain a variety of evolutionary phases of galaxy systems 
and their environment is suitable for studies of the properties of cosmic 
structures and the properties of galaxies therein in a wide range of densities 
in a consistent way, helping us to understand the role of environment in the 
evolution of galaxies and groups of galaxies. A number of studies have already 
shown that the supercluster environment affects the properties of 
galaxies, groups and 
clusters located there \citep{e2003b, 2004ogci.conf...19P, 
2005A&A...443..435W, 2006MNRAS.371...55H, e07b, 2008MNRAS.388.1152P,
2009arXiv0909.0232F, tempel09, Lietzen2009}.  

A comparison of the properties of rich and poor superclusters
have revealed a number of differences between them. 
The mean and maximum densities of galaxies in rich superclusters are larger than in poor
superclusters. Rich superclusters are more asymmetrical than poor superclusters
\citep{2007A&A...462..397E}. 
Rich superclusters contain 
high density cores  \citep{e07b,e08}. 
\citet{1998ApJ...492...45S} showed 
in a  study of the Corona Borealis supercluster that the core
of this supercluster may be collapsing.
The fraction of X-ray clusters in rich superclusters is 
larger than in poor superclusters \citep[][hereafter E01]{e2001}. The core 
regions of the richest superclusters may contain merging X-ray clusters 
\citep{2000MNRAS.312..540B, rose02}. The morphology of rich and poor superclusters 
is different---poor superclusters
(as an example we can use the Local Supercluster) resemble a ``spider''--a 
rich cluster with surrounding filaments while rich
superclusters resemble a ``multispider''---several high density regions (cores)
connected by filaments \citep{e07a}.

The extreme cases of observed objects usually provide the most stringent tests 
for theories; this motivates the need for a detailed understanding of the 
richest superclusters. So far, even their existence is not well explained by the
main contemporary structure modeling tool, numerical simulations:
while the masses of the richest simulated  superclusters are similar
to those of the richest observed superclusters \citep{e07a,2007A&A...462..397E,
ara08}, the number density of the richest superclusters
in simulations is about ten times smaller than the number density of
observed superclusters \citep{e06}. 
\citet{2010arXiv1003.4259Y} reached similar conclusions analysing  rich
systems in the 2dFGRS data.
\citet{2004ogci.conf...71B} mentioned
that the mass of the Shapley supercluster is very rarely reproduced in the
$\lambda$CDM model. 
It has even been argued that
the existence of very large superclusters contradicts
with the concept of a homogeneous Universe at large scales \citep{2009A&A...508...17S}.

The richest relatively close superclusters, which have been studied in detail,
are the Shapley Supercluster
\citep[][and references therein]{proust06} and the Horologium--Reticulum
Supercluster \citep{rose02, 2005AJ....130..957F, 2006AJ....131.1280F,
2009arXiv0909.0232F}.

Among the richest known galaxy systems, the Sloan Great Wall (SGW) deserves a 
special attention. The SGW is the richest and largest system of galaxies 
observed in the nearby Universe so far \citep{vogeley04,gott05,nichol06}. The 
SGW consists of several superclusters of galaxies; the richest of them is the 
supercluster SCl~126 from the E01 catalogue. The core of the supercluster SCl~126 
contains many rich clusters of galaxies, with X-ray clusters among them 
\citep{bel04, e07a}. Interestingly, the SGW has not been studied in detail yet, 
but already now we know that numerical simulations have been unable to reproduce 
several properties of the SGW. The SGW affects the measurements of the genus and 
Minkowski functionals of the SDSS and 2dF redshift surveys 
\citep{park05, saar06, gott08}, causing a ``meatball'' shift of the genus curve (a 
cluster-dominated morphology) of these surveys compared with the genus 
curve from simulations. \citet{croton04, nichol06} showed that the higher order 
moments of the correlation function  of the SDSS and 2dFGRS are sensitive to the 
presence of the SGW and cause a disagreement between higher order correlation 
functions in observations and in numerical simulations.  
\citep[Interestingly, the observed 
correlations can be explained by non-Gaussian initial density 
fields,][]{1996ApJ...462L...1G}. 
\citet{nichol06} mentions also that this disagreement may be due 
to the unusual morphology of the SGW. As we already mentioned, cosmological 
simulations have been unable to reproduce very rich extended systems as the SGW 
\citep{2009A&A...508...17S}.

\citet{e07a, e08} showed that the morphology of the richest supercluster in the 
SGW, the supercluster SCl~126 is unusual---it resembles a very rich filament 
with a high density core, while other rich superclusters (both  observed and 
simulated), the morphology of which we have studied, can in general be called 
``multispiders''---several high-density clusters are connected by lower density 
filaments.   This is another disagreement between the simulations and one 
supercluster in the SGW. 

\begin{figure*}[ht]
\centering
\resizebox{0.90\textwidth}{!}{\includegraphics*{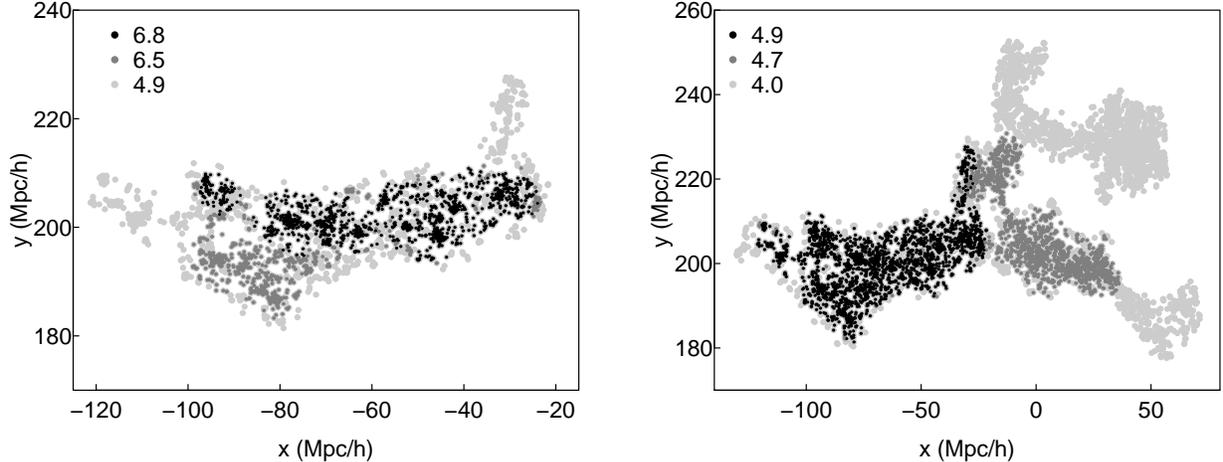}}
\caption{Distribution of galaxies in the Sloan Great Wall region. The coordinates x
  and y are in Mpc/$h$ (their definition is given in the Appendix, Eq.~\ref{eq.xyz}).
  The left panel shows the SCl~126 region, the right panel---a wider region
  with nearby superclusters. The surrounding region is shown in
  Fig.~\ref{fig:xy}. The relative luminosity density $D$ (normalized
  by the mean density) limits are as follows: left panel: black: $D\ge
  6.8$, dark grey: $D\ge 6.5$, light grey: $D\ge 4.9$.  Right panel:
  black: $D\ge 4.9$ (the same as the lightest level in the left panel),
  dark grey: $D\ge 4.7$, light grey: $D\ge 4.0$.  }
\label{fig:DF}
\end{figure*}

All this shows the importance of the studies of the morphology of superclusters. 
Information about the complex patterns of the large scale structure (the cosmic web) gets 
lost when we use, for example, correlation functions to study the clustering of 
galaxies \citep[see][for details and references]{2009LNP...665..493C} and also 
\citet{2009LNP...665..291V} for further reading about cosmic patterns). 
As we saw above, information about the morphology of the superclusters is useful 
when we interpret the results of the measurements of the correlation function. 

Morphology of superclusters can be used as one of the  tests to 
compare superclusters from observations and simulations. \citet{bas03} used 
supercluster shape as a cosmological probe to compare observed superclusters 
with those from simulations and showed that in this respect the $\Lambda$CDM model 
is in a better agreement with observations than the $\tau$CDM model.

In the present paper we study the morphology of the full SGW and of individual 
superclusters in the SGW with the aim to quantify the morphology of the SGW and 
to determine whether all superclusters in the SGW (and the full SGW) have 
peculiar morphologies not found in simulated superclusters. 
This study may help to answer the question whether the SGW is a genuine physical
structure \citep{2009LNP...665..291V}.
We use Minkowski 
functionals to compare substructure of superclusters in the SGW defined by 
different galaxy populations. We also compare the galaxy and group content of 
individual superclusters in the SGW with the goal to search for possible 
differences between superclusters in this aspect. 

First we delineate the superclusters in the SGW region, calculating the 
luminosity density field of galaxies from the SDSS DR7. We study the morphology 
of the SGW and individual superclusters in it, using Minkowski functionals and 
shapefinders. Next we study the morphology of these superclusters as determined 
by individual galaxy populations---this analysis gives us information about the 
substructure of superclusters. We define galaxy populations by their color, and 
add a separate population of bright red galaxies (BRGs). We assess the 
statistical significance of our results using Monte Carlo simulations.

Then we compare the galaxy and group content of the two richest superclusters in 
the SGW and search for differences between these superclusters. We compare the 
properties of groups in them, study the  group membership of BRGs, and analyze 
colors and morphological types of BRGs and the distribution of groups with 
different morphological types of BRGs in superclusters. We compare the peculiar 
velocities of elliptical and spiral BRGs and the first ranked galaxies in 
groups. The peculiar velocities of the first ranked galaxies in groups can be 
used as indicators of the dynamical state of groups and clusters 
\citep{coziol09}. Comparison of the peculiar velocities of the first ranked 
galaxies in groups in different superclusters of the SGW gives us information 
about the dynamical state of groups.

The BRGs satisfy the criteria for nearby ($z < 0.15$, cut I) LRGs 
\citep{eisenstein01}. \citet{eisenstein01} warn that the sample of 
nearby LRGs may be contaminated by galaxies of lower luminosity not suitable for 
a volume-limited sample, thus we call them BRGs (the name could be misleading, 
as the LRGs were first called thus; unfortunately, we have not found a better 
name). These galaxies can also be spirals; we will carry out their morphological 
classification. Understanding the spatial distribution and properties of the 
LRGs is of great cosmological importance, and our study of their nearby 
population (BRGs) will help to specify their properties.

\section{Data}
\subsection{Galaxies, groups and superclusters}
\label{subsec:data}

{\scriptsize
\begin{table*}[ht]
\centering
\caption{Superclusters in the region of the Sloan Great Wall.}
\begin{tabular}{rrrrrrrrrr} 
\hline 
(1)&(2)&\multicolumn{1}{c}{(3)}&(4)& (5)&(6)&(7)& (8)& (9)& (10)\\      
\hline 
Nr. & ID &\multicolumn{1}{c}{ID}& $N_{\mbox{gal}}$ &$N_{\mbox{gr}}$& $d_{\mbox{peak}}$  & $V$& $L_{\mbox{tot}}$ & $D_{\mbox{peak}}$& Diam.\\
    &         &             &       &        &Mpc/$h$ &(Mpc/$h$)$^3$& $10^{10}h^{-2} L_{\sun}$& <dens>& Mpc/$h$\\
    \hline
 1  &111      & 184+003+007 &1515   & 208    &  230.3 & 10565 &  1105.8 &  14.16  &   56.9           \\
 2  &111      & 173+014+008 &1334   & 190    &  242.0 & 10193 &  1050.8 &  12.31  &   50.3           \\
 3  &126,136  & 202-001+008 &3162   & 425    &  255.6 & 24558 &  2539.0 &  12.92  &  107.8           \\
 4  & 88      & 152-000+009 & 487   &  65    &  285.1 &  4666 &   453.5 &   9.81  &   39.0           \\
 5  &         & 187+008+008 & 711   &  73    &  267.4 &  4665 &   431.6 &   9.33  &   54.2           \\
 6  &         & 170+000+010 & 190   &  24    &  302.1 &  1841 &   171.0 &   8.55  &   20.1           \\
 7  &         & 175+005+009 & 348   &  51    &  291.0 &  3406 &   311.1 &   8.23  &   28.0           \\
 8  & 91      & 159+004+006 & 215   &  23    &  206.6 &  1108 &    98.3 &   7.61  &   15.4           \\
 9  & 91      & 168+002+007 & 387   &  34    &  227.7 &  1802 &   156.4 &   7.49  &   28.1           \\
10  &         & 214+001+005 & 332   &  33    &  162.6 &  1086 &    94.2 &   7.22  &   19.5           \\
11  &         & 189+003+008 & 468   &  70    &  254.1 &  2686 &   226.3 &   6.75  &   32.2           \\
12  &         & 198+007+009 & 100   &  18    &  276.0 &   649 &    53.7 &   6.33  &   13.1           \\
13  & 91      & 157+003+007 & 111   &   6    &  219.1 &   148 &    11.4 &   5.28  &   13.3           \\
\label{tab:scldata}                                                                                    
\end{tabular}
\tablecomments{
Columns in the Table are as follows:
\noindent 1: supercluster number, 
2: ID in the E01 catalogue,  
3: new ID (AAA+BBB+ZZZ, AAA---R.A., +/-BBB---Dec., CCC---100$z$),
4: the number of galaxies, 
5: the number of the T10 catalogue groups, 
6: distance of the density maximum,
7: volume,
8: total luminosity, 
9: peak density (in units of the mean density), 
10: diameter (the maximum distance between galaxies).
}
\end{table*}
}

We used the MAIN galaxy sample and the SDSS-LRG sample from the
7th data release of the Sloan Digital Sky Survey \citep{ade08,abaz08}. 
From these data we chose a subsample of the SGW region:
$150^\circ \leq R.A. \leq 220^\circ$, $-4^\circ \leq \delta \leq 8^\circ$, within the
distance limits 150~\Mpc$\le D_{com} \le $ 300~\Mpc.
This region fully covers the SGW and its surroundings, 
including poor superclusters, which connect the SGW with other nearby rich 
superclusters. In this region there are 27113 galaxies from the MAIN SDSS galaxy 
sample, and 3168 BRGs, which have been selected from the SDSS database; they 
are nearby ($z < 0.15$)   LRGs \citep{eisenstein01}
(cut I LRGs).  

We found groups of galaxies on the basis of the MAIN galaxy sample, 
as described in detail in \citet{2010A&A...514A.102T}, hereafter T10,
and \citet{tago08}. We corrected the redshifts of 
galaxies for the motion relative to the CMB and computed  the co-moving 
distances \citep{mar03} of galaxies using the standard cosmological 
parameters: the Hubble parameter $H_0=100 h\,$km\,sec$^{-1}$Mpc$^{-1}$, the matter 
density $\Omega_m = 0.27$, and the dark energy density $\Omega_{\Lambda} 
= 0.73$. Our SGW region sample contains about 4000 groups of galaxies.

In T10 groups of galaxies have been defined applying the Friends-of-Friends 
cluster analysis method introduced by \citet{tg76,zes82,hg82}, and modified by 
us \citep{tago08}. In this algorithm galaxies are linked into systems using a 
variable linking length. By definition, a galaxy belongs to a group of galaxies 
if this galaxy has at least one group member galaxy closer than the linking 
length. To take into account selection effects when constructing a 
group catalogue from a flux-limited sample, T10 calibrated the scaling of the
linking length with distance, using nearby rich groups as a template.  
As a result, the maximum group sizes in the sky projection and 
the velocity dispersions of groups in our catalogue are similar at all distances. 

The T10 group catalogue is available in electronic form at
the CDS via anonymous ftp to cdsarc.u-strasbg.fr (130.79.128.5) or
via \texttt{http://cdsweb.u-strasbg.fr/cgi-bin/qcat?J/  
A+A/514/A102}.


\begin{figure}[ht]
\centering
\resizebox{0.45\textwidth}{!}{\includegraphics*{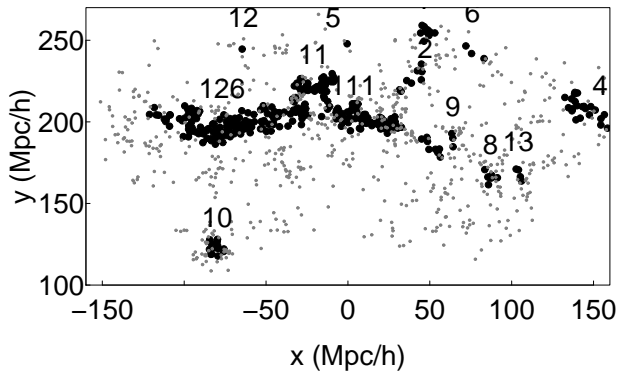}}
\caption{
Groups with at least four member galaxies in the Sloan Great Wall region
(for the SGW proper see Fig.~\ref{fig:DF}). 
Filled black circles show the groups in superclusters, 
and filled grey circles show those groups, which do not belong to superclusters.  
The numbers are the supercluster numbers in Table~\ref{tab:scldata}
(for the superclusters SCl~126 and SCl~111 we give the ID from the E01
catalogue, see text). 
}
\label{fig:xy}
\end{figure}

Next we calculated the luminosity density field of galaxies and found 
extended systems of galaxies (superclusters) in this density field. 
This procedure is described in more detail in Appendix~\ref{sec:DF}. 
Supercluster catalogue is described in detail in
\citet{2010arXiv1012.1989J}.

To search for systems (high density regions) in the distribution of galaxies and 
clusters of galaxies usually either the Friend-of-Friend algorithm
or a density field approach is used. In the Friend-of-Friend method galaxies or
clusters of galaxies are collected together so that each 
object has at least one neighbor closer than the linking length
(a fixed distance). For small linking lengths
all objects are isolated. When the linking length increases more and more objects
are collected into systems, and as a result, the richnesses and lengths
of systems increase. At a certain linking length systems join rapidly
and form a system that may penetrate the whole volume under study---percolation
occurs. Superclusters as the largest still relatively isolated 
systems are determined using a linking length smaller than the 
percolation length \citep[for details we refer to][]
{1993ApJ...407..470Z,1994MNRAS.269..301E,e2001}. 
When using the density field to select systems,
we search for connected systems with the density higher than a certain
threshold. When we decrease the density threshold, then, as with increasing 
linking length in the Friend-of-Friend method, 
more objects join the systems, and the richnesses and lengths of systems increase.
At a certain threshold the systems percolate the sample volume. 
Again, supercluster catalogues are compiled using
a density threshold higher than the percolation threshold
\citep{2003A&A...410..425E,2003A&A...405..425E,2004MNRAS.352..939E,
2007A&A...462..811E}.
\citet{2004MNRAS.352..939E} used a varying
density threshold to determine superclusters using the 2dF data; for details 
we refer to their study. 
While the Friend-of-Friend method tends to find elongated systems
\citep{1984MNRAS.206..529E},
the density field approach does not have such a bias.

Comparison of different catalogues of superclusters shows that rich superclusters
can be identified in different ways, with similar parameters (e.g., the maximum
length). For example, the supercluster SCl~126 has been 
identified in \citet{2003A&A...405..425E}  using the SDSS data and the density
field (the supercluster N13 in the list), and has a length of 90\Mpc;
in \citet{2003A&A...410..425E} this supercluster is determined using the 2dF data,
this is the supercluster SCl~152 in the list with the length of 112\Mpc.
In \citet{2004MNRAS.352..939E} list this supercluster is the supercluster SCNGP06.

In order to choose proper density levels to determine the SGW and the individual 
superclusters, which belong to the SGW, we compared the  density field 
superclusters at a series of density levels (Fig.~\ref{fig:DF}). We analyzed the 
systems of galaxies at each level, and examined the joining events, when we 
moved from higher density levels to lower ones. We will use the relative 
luminosity density $D$ here and below (normalized to the mean density in the 
SDSS sample volume). At a very high density level ($D = 6.8$) only the highest 
density part of the SGW is seen. This can be identified with the core region of 
the supercluster SCl~126 in the E01 catalogue. We use this density level to 
define the core of this supercluster, and denote it as SCl~126c. As we decrease 
the density level, galaxies from the outskirts of the SCl~126 join the SGW. 
At the density level $D = 4.9$ the richest two superclusters in the 
SGW (SCl~126 and SCl~111) 
still form separate systems. At a lower density level, $D = 4.7$, these two 
superclusters join together to become one connected 
system, and several other superclusters also join this system. At a still lower 
density level ($D = 4.0$) these superclusters, which border the voids around the 
SGW,  join the SGW. Several poor superclusters, which form a low-density 
extension of the SGW, join the SGW at this level, too.
We show in Fig.~\ref{fig:DF} the full SGW region 
and the surrounding superclusters at different  density levels
from $D = 6.8$ to  $D = 4.0$.

We used the information about joining of superclusters to 
determine the SGW and the individual superclusters in the
SGW as follows. 

First, we used the density level $D = 4.9$ to determine individual superclusters, 
which belong to the SGW. In Fig.~\ref{fig:xy} we show and mark all superclusters 
of galaxies in the SGW region. The data on these superclusters are 
presented in Table~\ref{tab:scldata}.

The richest superclusters in the SGW are the superclusters SCl~126 and 
the supercluster SCl~111. The supercluster SCl~126 includes also the 
supercluster SCl~136 from the E01 catalogue. We note that the supercluster 
SCl~111 in the E01 catalogue actually corresponds to two superclusters in 
the SGW region, the first and second supercluster in Table~\ref{tab:scldata}. 
In what follows we will call the first supercluster in this table SCl~111, 
as the second supercluster  actually 
does not belong to the SGW (see Fig.~\ref{fig:xy}). We will use the notation SCl~126c 
for the core region of the supercluster SCl~126, and the notation SCl~126o 
for the outskirts of this supercluster, defined by excluding the 
core SCl~126c from the whole supercluster SCl~126. 

We use the density level $D = 4.7$ to determine the whole SGW. This adds to 
the SGW also those poor superclusters, which form the low density extension of 
the SGW (superclusters 9, 8,  and 13 in Table~\ref{tab:scldata}, which all belong 
to the SCl~91 in the E01 catalogue). Thus, the highest density part of the SGW is the 
supercluster SCl~126 and the lowest density part is formed by poor 
superclusters, members of the supercluster SCl~91 in the E01 
catalogue. In addition, we combine those galaxies, groups and poor superclusters 
in the volume defined at the start of this Section, which do not belong to the SGW,
into a comparison 
sample and call it the field sample (denoted as F).
Our SGW sample contains 6138 galaxies and 817 groups, 
351 of which host BRGs. The field sample contains 20975 galaxies and
3020 groups; 1589 groups in the field have BRGs in them. In the SGW
83\% of all BRGs belong to groups, in the field---68\%.

\subsection{Galaxy populations}
\label{subsec:pops}

\begin{figure}[ht]
\centering
\resizebox{0.48\textwidth}{!}{\includegraphics*{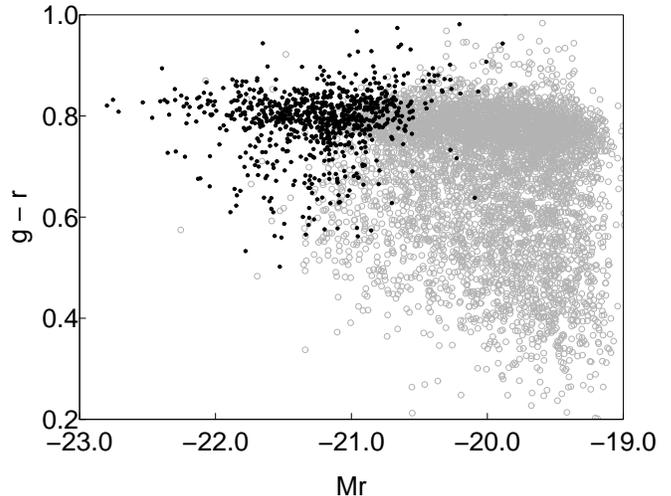}}
\hspace*{2mm}\\
\caption{The color-magnitude diagram of the SGW galaxies. Open circles -- main sample galaxies, filled circles -- BRGs.
}
\label{fig:mainvsBRG}
\end{figure}

In the present paper we use the $g - r$ colors of galaxies and their absolute 
magnitude in the $r$-band $M_r$. The absolute magnitudes of individual galaxies 
are computed using the the KCORRECT algorithm 
\citep{blanton03a,2007AJ....133..734B}. We also 
accepted $M_{\odot} = 4.53$ (in the $r$ photometric system). Evolution 
correction $e$ has been calibrated according to \citet{blanton03b}. All of our 
magnitudes and colors correspond to  the rest-frame at the redshift $z=0$. We 
note that our colors are found using the rest-frame Petrosian magnitudes, not 
the model magnitudes as done by \citet{eisenstein01}, so there might be small 
differences in the colors.

\begin{figure*}[ht]
\centering
{\resizebox{0.32\textwidth}{!}{\includegraphics*{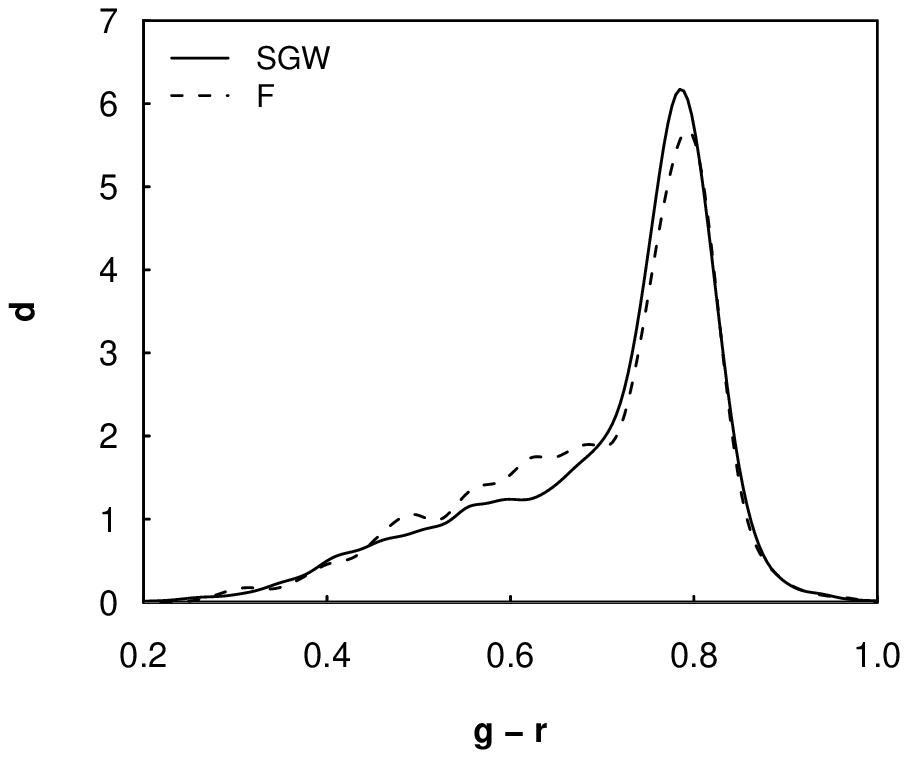}}}
{\resizebox{0.32\textwidth}{!}{\includegraphics*{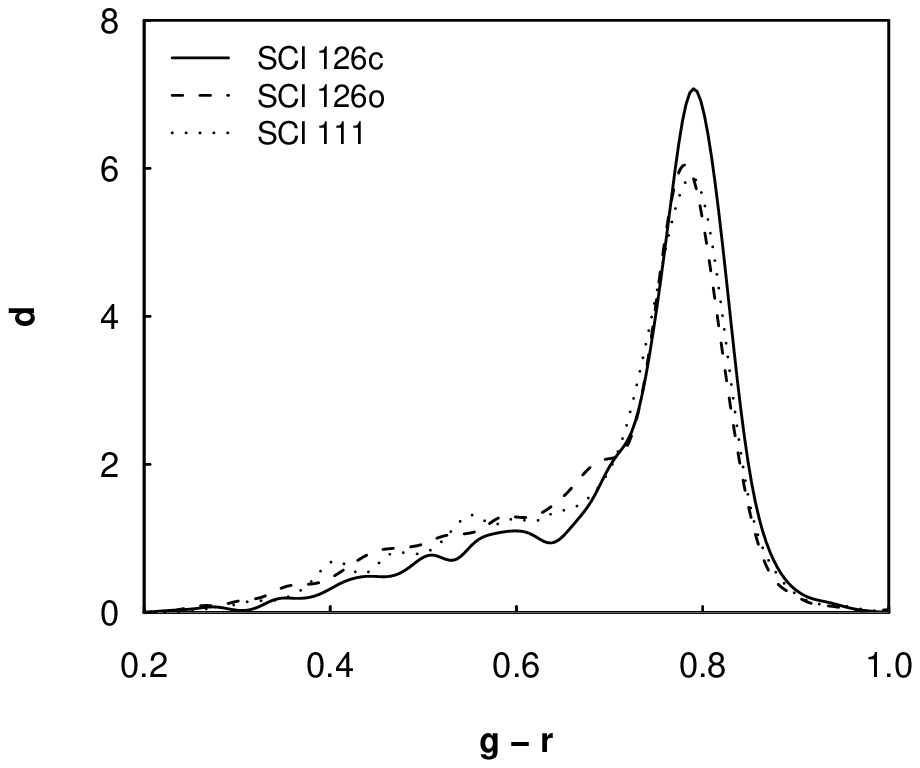}}}
{\resizebox{0.32\textwidth}{!}{\includegraphics*{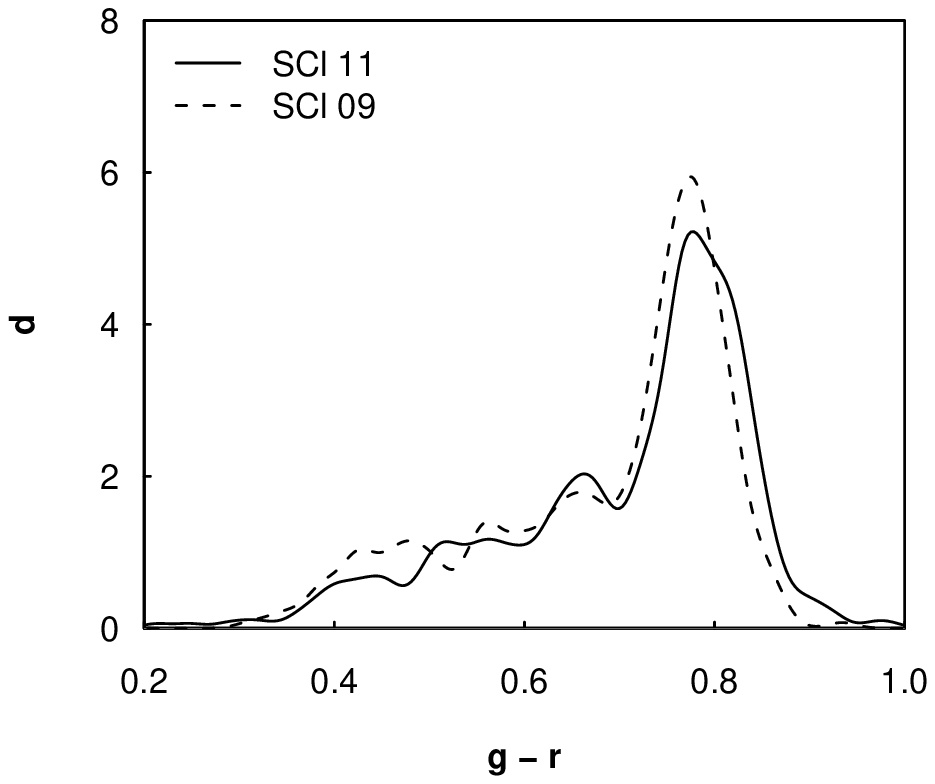}}}
\hspace*{2mm}\\
\caption{
The distribution of  the $g - r$  colors for galaxies in the SGW and in
the field (F) (left panel), in the supercluster 
SCl~126 (core and outskirts, SCl~126c and SCl~126o) and SCl~111 (middle panel), 
and  in the superclusters
9 and 11 (right panel, see Table~\ref{tab:scldata} for supercluster ID).
The color distributions were calculated for the absolute magnitude $M_r$ 
limit $M_{r0}\le -19.25$. All distributions were calculated using the 'density' command
in the R statistical environment with the bandwidth 0.015.
The numbers of galaxies in each sample are as follows:
SGW---5925, field---18611, SCl~126c---1397, SCl~126o---1751,
SCl~111---1419, SCl~09---355, and SCl~111---454 galaxies.
}
\label{fig:galpop}
\end{figure*}

As said above, we will pay special attention to two populations of galaxies, the 
first-ranked galaxies of groups, and the BRGs (the galaxies that have the 
GALAXY\_RED flag in the SDSS database). The BRGs are nearby 
($z < 0.15$, cut I) LRGs \citep{eisenstein01} and are similar to the LRGs 
at higher redshifts. Nearby bright red galaxies 
do not form an approximately volume-limited 
population, but they are yet the most bright and the most red galaxies in the SGW 
region, as shown in Fig.~\ref{fig:mainvsBRG}. 

BRGs have been selected using model magnitudes, but in our analysis we will use 
rest-frame Petrosian magnitudes for all galaxies including BRGs, for concordance.

For a small subsample of galaxies (the BRGs and the first-ranked galaxies in groups,
Sect.~4) 
we determined also the morphological types of galaxies, using data from the SDSS 
visual database. We used pseudo true-color images and
followed the classification, used in the Galaxy Zoo project 
\citep{lintott08}---elliptical galaxies are of type 1, spiral galaxies 
of type 4, those 
galaxies which show signs of merging or other disturbancies, are of type 6. 
Types 2 and 3 denote clockwise and anti-clockwise spiral galaxies, for which spiral 
arms are visible. Fainter galaxies, for which the morphological type is difficult 
to determine, are of type 5. The use of pseudo-color images may, in principle,
intruduce a bias toward blue-looking galaxies to be classified 
as spirals and red-looking galaxies---as ellipticals. We tried to avoid this:
several people classified galaxies independently. Later comparison
showed that in most cases 
these independent classifications coincided. Our sample of the BRGs and the 1st ranked galaxies 
consists of quite bright galaxies, so it was possible to determine 
their morphological type reliably. 
 
\citet{eisenstein01} warn that the sample of nearby LRGs 
may be contaminated by galaxies of lower luminosity not suitable for the use of 
nearby BRGs as a volume-limited sample. Nearby BRGs may also be of late type and/or,
as shows  
Fig.~\ref{fig:mainvsBRG}, some of them may have blue $g - r$ colors. Later 
we will compare the colors and morphological types of BRGs in color-magnitude 
diagrams to check for probable presence of blue, late type galaxies in our 
sample of BRGs.

Images of individual galaxies, which we used for morphological classification,
can be found at
\url{\texttt{http://www.aai.ee/$\sim$maret/SGWmorph.html}}.

We divide the galaxies into the red and blue populations, using the color 
limit $g - r \geq 0.7$  (we call these galaxies red). We use volume
limited samples with the $r$-band magnitude limit $M_r = -19.25$. We show in 
Fig.~\ref{fig:galpop} the distributions of the $g - r$ colors of galaxies in the 
whole SGW and in the individual superclusters in the SGW. The numbers of galaxies 
in each sample are given in caption. Comparison with the numbers of galaxies in 
superclusters in Table~\ref{tab:scldata} shows that actually all samples
are almost volume limited---the selection effects were
taken into account correctly when we calculated the luminosity density field.

In the present paper  all the distributions (probability densities)  were 
calculated within the R environment using the 'density' command in the 'stats' 
package \citep{ig96}, \texttt{http://www.r-project.org}. The 'density' function 
chooses the optimal bin width minimizing the MISE (mean 
integral standard error) of the estimate. This package does not provide the 
customary error limits (local error estimates), but the usual Poisson errors
do not describe these, also \citep[we explained this in more detail in][]{e08}.
We compare the density 
distributions using the full data (i.e. the integral distributions) and 
statistical tests, as the Kolmogorov-Smirnov test, throughout 
the paper.

Fig.~\ref{fig:galpop} shows 
that in the whole SGW the fraction of red galaxies is larger than in the field, 
an evidence of the large-scale morphological segregation \citep{e07b}. In the 
SGW, 65\% of galaxies are red, in the field---51\%. In addition, there are 
relatively more red galaxies in the core of the supercluster  SCl~126 (SCl~126c) 
than in other superclusters (Table~\ref{tab:sclgal}). This agrees with earlier 
findings by \citet{e07a,e07b}. We used the Kolmogorov-Smirnov test to estimate the
statistical significance of the differences between the colors of galaxies in 
different superclusters. This test showed that the differences between the colors 
of galaxies in the SGW and in the field have a very high significance (the 
probability that the color distributions were drawn from the same sample is less 
than $10^{-5}$). The differences between the colors of galaxies in the core of the 
supercluster SCl~126c and in the outskirts of this supercluster, SCl~126o, and 
between the colors of galaxies in  the SCl~126c and the colors of galaxies in other 
superclusters, have also a very high statistical significance. The differences 
between the colors of galaxies in other superclusters are marginal.

\begin{figure*}[ht]
\centering
{\resizebox{0.95\textwidth}{!}{\includegraphics*{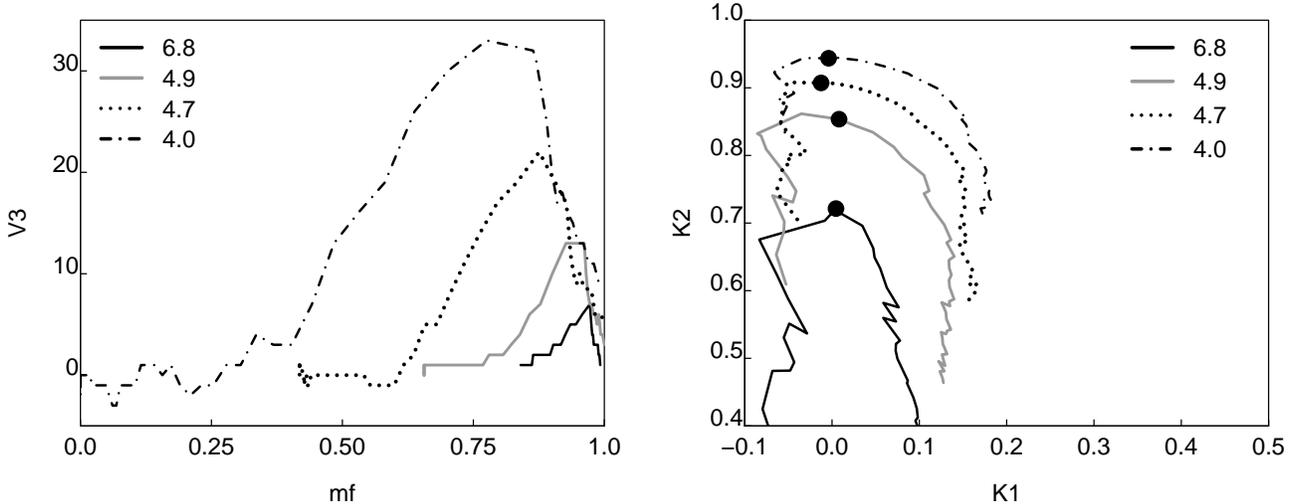}}}
\hspace*{2mm}\\
\caption{
The 4th Minkowski functional $V_3$ and the shapefinders 
 $K_1$ (planarity)
 and $K_2$ (filamentarity) for the SGW for different density levels. 
The mass fraction in the left panel is that 
for the whole SGW at the density level $D = 4.0$; 
for other density levels the superclusters start at higher mass fractions
(sect. 3.3). 
In the right panel the mass fraction $m_f$ increases anti-clockwise along the 
curves.
 The filled circles in the right panel mark the value of the mass fraction 
 $m_f=0.7$. In the $(K_1,K_2)$-plane filaments are located near the $K_2$-axis,
pancakes near the $K_1$-axis, and ribbons along the diagonal.
Solid black line corresponds to the density level $D = 6.8$,
solid grey line---$D = 4.9$, dotted line---$D = 4.7$, and dash-dotted line---$D = 4.0$. 
}
\label{fig:MF}
\end{figure*}

In the T10 catalogue the first ranked galaxy of a group 
is defined as the most luminous galaxy in the $r$-band. We will use the same definition 
in the present paper; later we will compare the properties of those first rank 
galaxies, which are BRGs, with those, which are not BRGs. 

A 3D interactive PDF graph that shows the spatial distribution
of groups of galaxies in the SGW can be found at
\url{\texttt{http://www.aai.ee/$\sim$maret/SGWmorph.html}}.

\section{Morphology}
\subsection{Minkowski functionals and shapefinders}

If we want to understand the formation, evolution and present-day properties
of cosmic structures then we must use, in addition to the usual statistics
that characterize the clustering of galaxies 
\citep[the 2-point correlation function,
Nth nearest neighbour and others, see][for a review about methods
to characterize galaxy distribution]{mar03} 
higher order statistics
to describe structures like superclusters, filaments and clusters
in which galaxies are embedded \citep{2010MNRAS.406.1609B}.
A number of methods are used to define and characterize cosmic
structures 
\citep[][and references therein]{2005PASA...22..136P, 2010A&A...510A..38S,
2009arXiv0912.3448V, 2010ApJ...723..364A,
2010MNRAS.406.1609B, 2010MNRAS.409..156B}. 
One possible approach is to determine cosmic structures (in our case---superclusters
of galaxies) using, for example, a density field, and then
to study their morphology 
\citep[][and references therein]{1997ApJ...482L...1S,1998ApJ...508..551S, sss04}.
    
Supercluster geometry (morphology) is defined by its outer
(limiting) isodensity surface, and its enclosed volume. When
increasing the density level over the threshold overdensity
$D = 4.7$ (Sect.~\ref{subsec:data}), the isodensity surfaces move into the
central parts of the supercluster.  The morphology of the
isodensity contours is (in the sense of global geometry) completely
characterized by the four Minkowski functionals \mbox{$V_0$--$V_3$}
 (we give the formulas in the Appendix \ref{sec:MF}).

Minkowski functionals, genus, and shapefinders 
have been used earlier to study the morphology of the large scale 
structure in the 2dF and SDSS surveys \citep{hik03, park05, saar06, jam06, gott08} and 
to characterize the morphology of superclusters \citep{sah98,
bpr01, kbp02, 2003MNRAS.343...22S, 
bas03, sss04, bas06} in observations and simulations. These studies 
concern only the ``outer'' shapes of superclusters and do not treat their 
substructure. We expanded this approach in \citet{e07a}, using the Minkowski 
functionals and shapefinders to analyze the full density distribution in 
superclusters, at all density levels.

The original luminosity density field, used to determine individual 
superclusters in the SGW and the full SGW complex, was calculated using all galaxies. 
To calculate the density field for the study of the 
morphology, we recalculated the density field for each individual structure under 
study. We used volume-limited galaxy samples and calculated the number density; 
this approach makes our results 
insensitive to selection corrections (although our
SGW samples are almost volume-limited already, compare the numbers of galaxies
in full and volume-limited samples, given in Table~\ref{tab:scldata}
and in the caption of Fig.~\ref{fig:galpop}).  To obtain the density field for
estimating the Minkowski functionals, we used a kernel estimator with a $B_3$ 
box spline as the smoothing kernel, with the radius of 16~\Mpc\ 
\citep{saar06,e07a}. As the argument labeling the isodensity surfaces, we chose 
the (excluded)  mass fraction $m_f$---the ratio of the mass in regions with 
density {\em lower} than the density at the surface, to the total mass of the 
supercluster. 
When this ratio runs from 0 to 1, the iso-surfaces move from the 
outer limiting boundary into the center of the supercluster, i.e. the fraction 
$m_f=0$ corresponds to the whole supercluster, and $m_f=1$ to its highest 
density peak. This is the convention adopted in all papers devoted to the 
morphology of the large-scale galaxy distribution. 
The reason for this convention (the higher
the density level, the higher the value of the mass fraction) is 
historical---the most popular argument for the genus and for the
Minkowski functionals has been the
volume fraction that grows with the density level; all other arguments
are chosen to run in the same direction. We refer to \citet{e07a} for details
of the calculations of the density field.

For a given surface the four Minkowski functionals (from the first to the
fourth) are proportional to: the enclosed volume $V$, the area of the surface
$S$, the integrated mean curvature $C$, and the integrated Gaussian curvature
$\chi$. 
Last of them, the fourth Minkowski functional $V_3$ describes the
topology of the surface; at high densities it gives the number of isolated
clumps (balls) in the region, at low densities -- the number of cavities
(voids) \citep[see, e.g.][]{saar06}. 

The fourth Minkowski functional $V_3$ is well suited 
to describe the clumpiness of the galaxy distribution inside 
superclusters---the fine structure of superclusters. We calculate 
$V_3$ for galaxies of different populations in 
superclusters for a range of threshold densities, starting with the lowest 
density used to determine superclusters, up to the peak density in the 
supercluster core. So we can see in detail how 
the morphology of superclusters is traced by galaxies of different type.

In addition to the fourth Minkowski functional $V_3$
we use the 
dimensionless shapefinders $K_1$ (planarity) and $K_2$ (filamentarity)
\citep{sah98} 
(Appendix \ref{sec:MF}).  In \citet{e07a} we showed that in the 
$(K_1,K_2)$ shapefinder plane the morphology of superclusters is described 
by a curve (morphological signature) that is similar for all multi-
branching filaments. In the present paper we will use the morphological 
signature to characterize the morphology of superclusters and of 
galaxy populations in them.

There are two main effects that affect the reconstructed density field and its 
Minkowski functionals. The first is due to the fact that about 6-7\% of galaxy 
redshifts are missing, because of fiber collisions \citep{2010arXiv1005.2413T}. 
About 60\% of these are 
approximately at the same distance as their close neighbours 
\citep{zehavi02}, 
but others are not. We found the collision groups in the SDSS, determined the 
galaxies that do not have redshifts because of fiber collisions, and ran Monte-
Carlo simulations, assigning to 60\% of randomly selected non-redshift galaxies 
the redshift of their neighbours. Other 40\% of non-redshift galaxies were 
omitted, assuming that their true redshift will take them out of the 
supercluster. This procedure should give upper limits for errors. We ran 1000 
simulations and found that both the morphological signatures and $V_3$ did not 
change (at the 95\% confidence level) -- fiber collisions are too scarce to 
affect the estimates of the morphological characteristics.

Another effect is the discreteness of galaxy catalogues (shot noise) that 
induces errors in the density estimates. To take this into account we have to 
define first the statistical model for the spatial distribution of galaxies 
within a supercluster. One possibility is to use the Cox model, where we have a 
realization of a random field and a Poisson point process populates the 
supercluster by dropping galaxies there with the intensity proportional to the 
value of the realization in the neighbourhood of a possible galaxy. We tested 
that model and found that it is not able to describe the structure of 
superclusters; we summarize the results in App.~\ref{sec:Cox}.

After several erroneous approaches, we used the popular halo model ideology 
\citep[see][and references therein for 
details]{cooray2002,2007ApJ...671..153Y,2007MNRAS.376..841V} 
to define a statistical model for superclusters. We assume that 
supercluster structure is defined by its dark matter haloes, and discreteness 
errors are caused by the random positions of galaxies within these haloes. As 
the halo model assumes that the main galaxy of a halo lies at its centre, the 
main galaxies of our groups and the isolated galaxies (the main galaxies of 
haloes where other galaxies are too faint to see) remain fixed. For satellite 
galaxies, we use smoothed bootstrap to simulate the distribution of satellites 
inside our haloes -- we select satellite galaxies by replacement, and add to 
their spatial positions increments with the same distribution as the $B_3$ 
kernel we use for density estimation. Smoothed bootstrap is known to be able to 
estimate pointwise errors of densities \citep{silverman87,davison2009}. As the 
the $B_3$ spline (Eq.\ref{eq:b3}) practically coincides with a Gaussian of 
the rms deviation $\sigma=0.6$, we define the $B_3$ kernel width $a=\sigma_r/0.6$ for every halo (galaxy 
group) using the rms deviation $\sigma_r$ of the radial positions of galaxies for that group. 
We generated mock superclusters with 
new galaxy distributions by smoothed bootstrap and calculated again 
the density field and Minkowski functionals, repeating this procedure 1000 
times. The figures in Sect. 3.4, where we compare the Minkowski functionals and 
shapefinders for galaxy populations of different superclusters in the SGW show 
the 95\% confidence regions obtained by these simulations.

\subsection{Minkowski functionals and shapefinders---general}

Next we describe shortly how the 4th Minkowski functional $V_3$
and the shapefinders $K_1$ and $K_2$ describe the morphology of a system.

At small mass fractions the isosurface includes the whole supercluster and the 
value of the 4th Minkowski functional $V_3 = 1$.  As we move to higher mass 
fractions, the isosurface includes only the higher density parts of 
superclusters. Individual high density regions in a supercluster, which at low 
mass fractions are joined together into one system, begin to separate from each 
other, and the value of the 4th Minkowski functional ($V_3$) increases. At a 
certain density contrast (mass fraction) $V_3$ has a maximum, showing the 
largest number of isolated clumps in a given supercluster. 
The 4th Minkowski functional $V_3$ depends also on the number
of tunnels, which may appear in the supercluster topology,
as part of the galaxies do not contribute to the supercluster
at intermediate density levels. At still higher 
mass fractions only the high density peaks remain in the supercluster and 
the value of $V_3$ decreases again.

When we increase the mass fraction, the changes in the morphological signature 
accompany the changes of the 4th Minkowski 
functional. As the mass fraction  increases, at first the planarity 
$K_1$ almost does not change, while the filamentarity $K_2$ increases
-- at higher density levels superclusters become more filament-like 
than the whole supercluster. Then also the planarity starts to decrease, and at a mass 
fraction about $m_f = 0.7$ the characteristic morphology of a 
supercluster changes -- we see the crossover from the outskirts of a supercluster 
to the core of a supercluster \citep{e07a}.

\subsection{Morphology of the SGW}

We calculate the Minkowski functionals and shapefinders for superclusters in the 
SGW, defined at four different density levels, to see the change of the morphology of the 
SGW with density (Fig.~\ref{fig:MF}).
In the left panel of this figure the individual mass fractions for higher density levels are rescaled
and shifted 
so that they show the fraction of mass of systems at a high density level relative
to the total mass of the systems at the lowest density level, $D = 4.0$.

\begin{figure*}[ht]
\centering
{\resizebox{0.95\textwidth}{!}{\includegraphics*{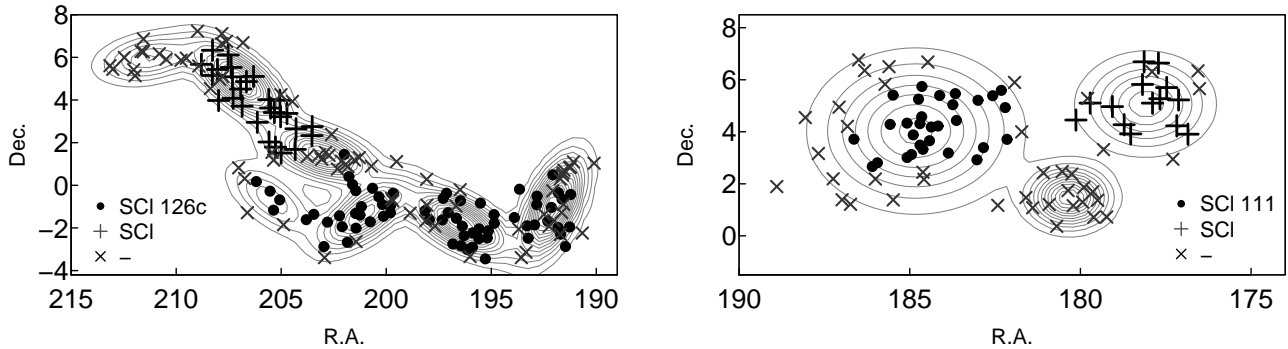}}}
\\
\caption{
The sky distribution of galaxies and BRGs in the superclusters SCl~126 and SCl~111.
The distribution of galaxies in the whole supercluster (at the density level 
$D = 4.9$) is shown by density 
contours. Dots mark groups with BRGs which belong to superclusters
at the density level $D = 6.8$, crosses mark groups with BRGs in other systems
at this density level, which have not
joined the superclusters SCl~126 and SCl~111 yet, and x-s mark groups with 
BRGs which at this density level do not belong to any system.  
}
\label{fig:sky68}
\end{figure*}

\begin{figure*}[ht]
\centering
\resizebox{0.53\textwidth}{!}{\includegraphics*{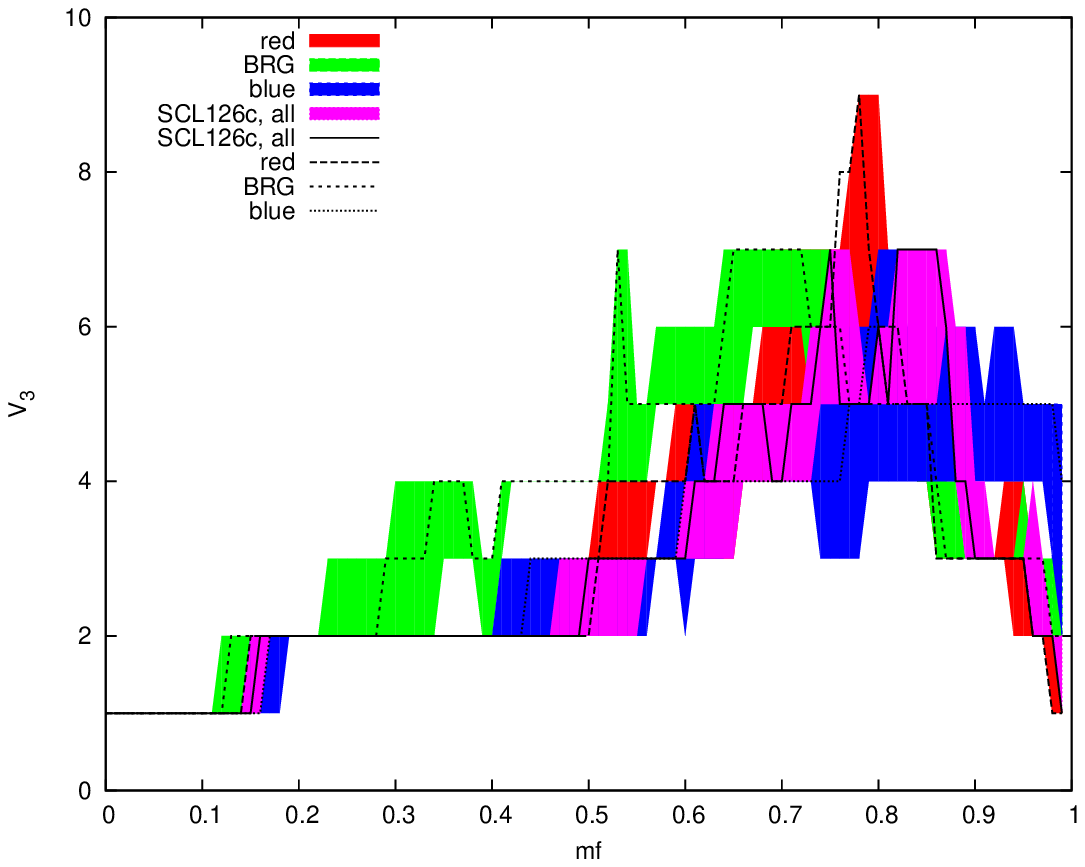}}
\resizebox{0.262\textwidth}{!}{\includegraphics*{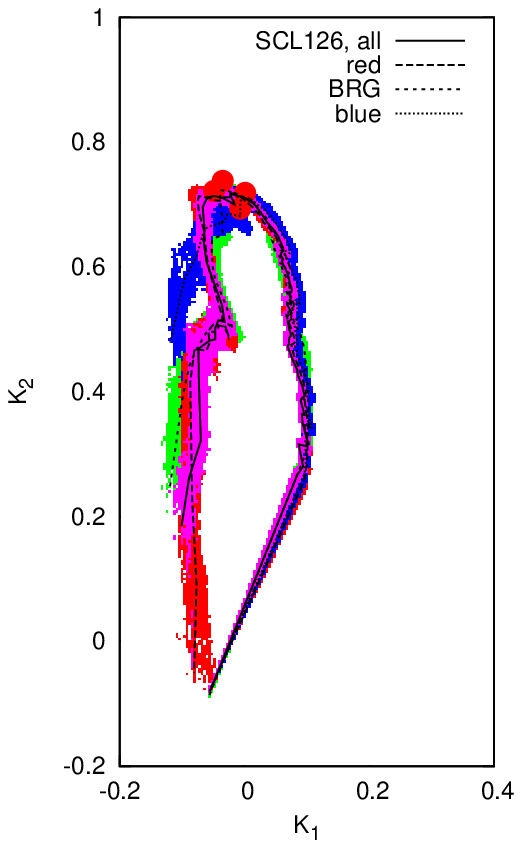}}
\\
\caption{
The 4th Minkowski functional $V_3$ (left panel) and the shapefinders 
 $K_1$ (planarity) and $K_2$ (filamentarity) 
 (right panel) for the red and blue galaxies and the BRGs in  
the core of the supercluster SCl~126 (SCl~126c). 
The black line denotes  $V_3$ for all galaxies (given for comparison,
this is the same line as in Fig.~\ref{fig:MF} for the density level
$D = 6.8$, but here the mass fraction is not rescaled), 
dashed line with long dashes---red galaxies, 
dashed line with short dashes---BRG-s, and dotted line---blue galaxies.
The filled circles in the right panel mark the value of the mass fraction 
$m_f\approx  0.7$. 
The colored regions in this and the next two figures
show the 95\% confidence regions
obtained by bootstrap simulations as explained in text. With red color 
we show the confidence regions for red galaxies, blue corresponds
to blue galaxies, green to BRGs and magenta to the full supercluster.
}
\label{fig:MFG1}
\end{figure*}

\begin{figure*}[ht]
\centering
\resizebox{0.53\textwidth}{!}{\includegraphics*{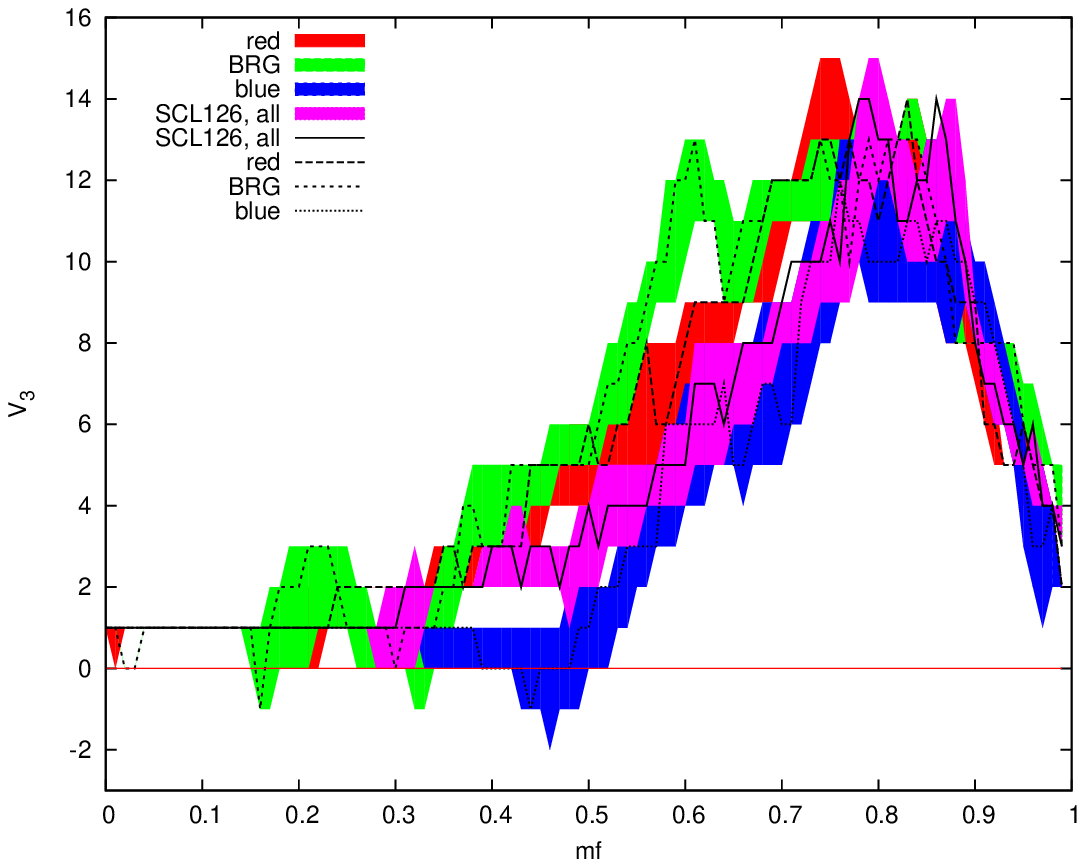}}
\resizebox{0.262\textwidth}{!}{\includegraphics*{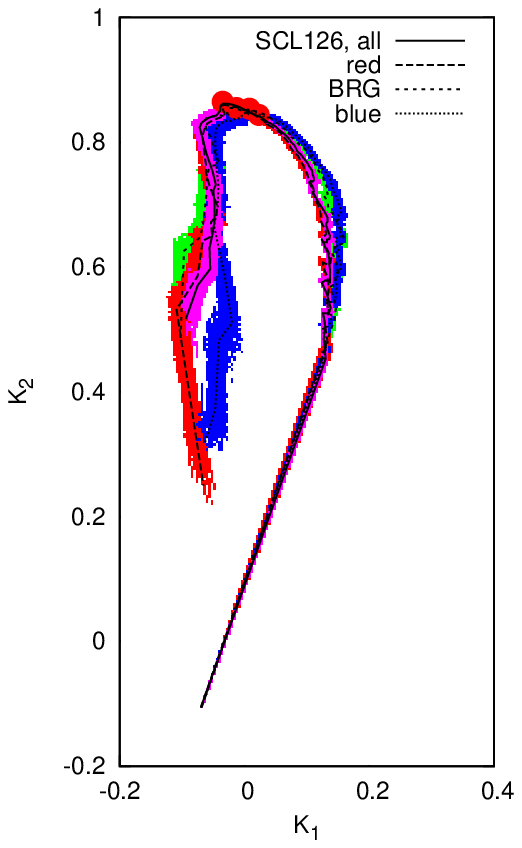}}
\\
\caption{
The 4th Minkowski functional $V_3$ and the shapefinders 
$K_1$-$K_2$ for the red and blue galaxies and the BRGs in  
the whole supercluster SCl~126. 
The lines and colored regions
correspond to galaxy populations 
as explained in Fig~\ref{fig:MFG1}. Again,
for comparison we show $V_3$ and $K_1$-$K_2$ lines for the full
supercluster.
The filled circles in the right panel mark the value of the mass fraction 
$m_f\approx 0.7$.
}
\label{fig:MFG2}
\end{figure*}

At the highest density level ($D = 6.8$) only the core region of the 
supercluster SCl~126 contributes to the SGW. The left panel of Fig.~\ref{fig:MF} 
shows that for this core, the value of the 4th Minkowski functional 
$V_3$ is small almost over the whole range of mass fractions,
the largest value of $V_3$ is 5. 
The right panel of Fig.~\ref{fig:MF} shows the corresponding 
morphological signature, with the starting planarity $K_1  \approx 0.1$. As 
the mass fraction increases, the filamentarity $K_2$ increases and 
reaches a maximum value of about 0.72 at the mass fraction $m_f = 0.7$, when 
the morphological signature changes. The shape of the morphological signature $K_1$-
$K_2$ is consistent with a single filament \citep[see for details ][]{e07a}.

\begin{figure*}[ht]
\centering
\resizebox{0.53\textwidth}{!}{\includegraphics*{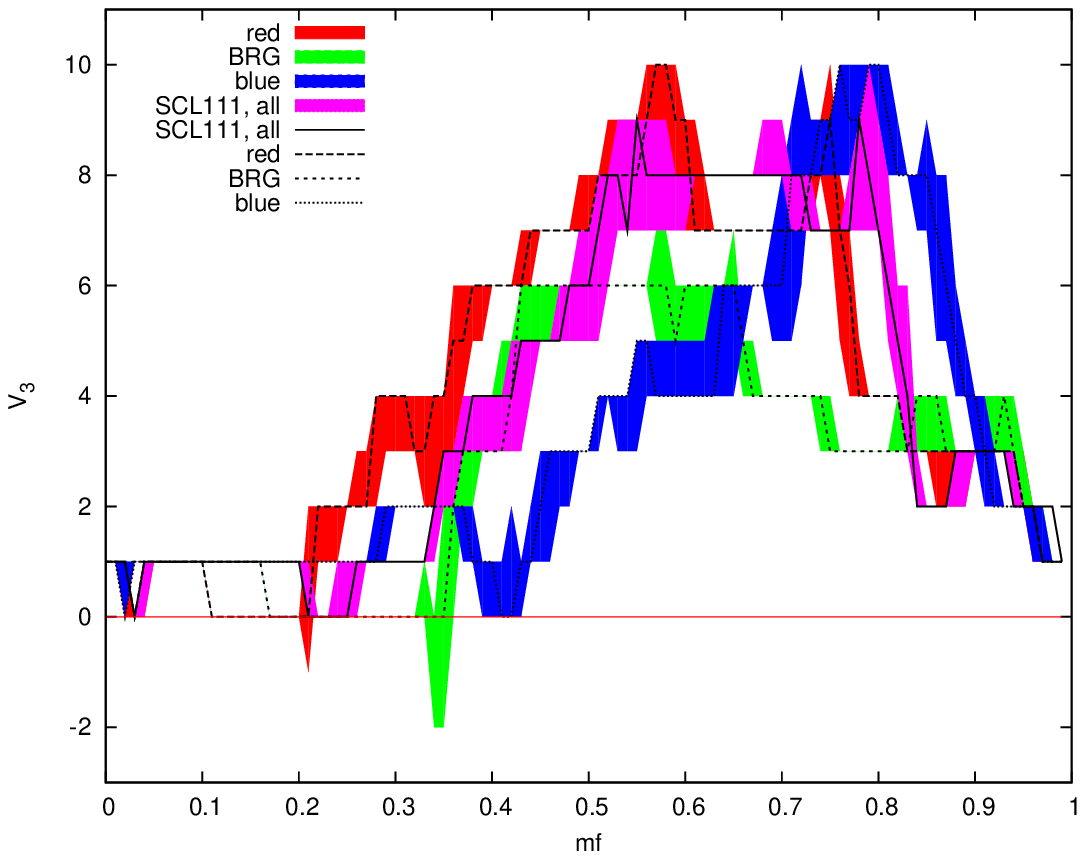}}
\resizebox{0.262\textwidth}{!}{\includegraphics*{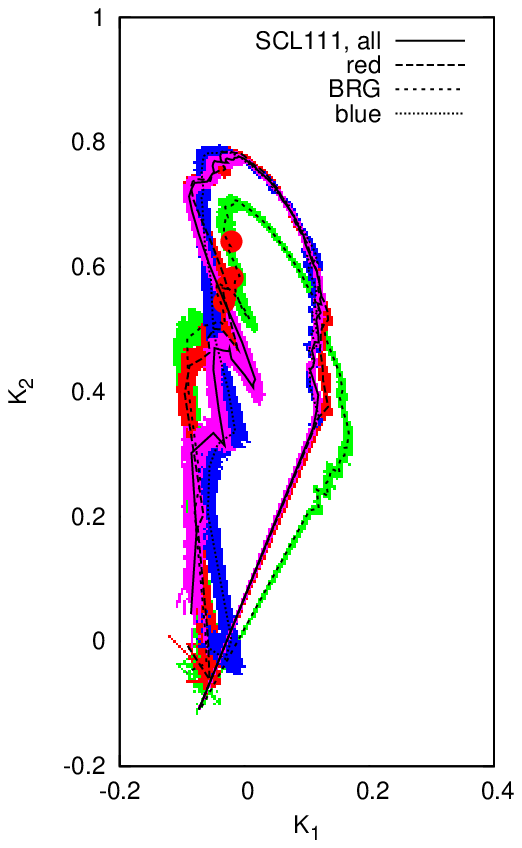}}
\\
\caption{
The 4th Minkowski functional $V_3$ and the shapefinders 
$K_1$- $K_2$ for the red and blue galaxies and the BRGs in  
the supercluster SCl~111. 
The lines and colored regions
correspond to galaxy populations 
as explained in Fig~\ref{fig:MFG1}.
The filled circles in the right panel mark the value of the mass fraction 
$m_f\approx 0.7$.
}
\label{fig:MFG3}
\end{figure*}

At the density level $D = 4.9$ new galaxies join the SGW and the SGW coincides 
with the whole supercluster SCl~126. The maximum value of the clumpiness $V_3 = 
12$, and the values of the planarity $K_1$ and filamentarity $K_2$  are larger 
-- the morphology of the supercluster resembles a multibranching filament 
\citep{e07a}, as shown by the morphological signature. 

At the density level $D = 4.7$ the supercluster SCl~111 also joins the 
SGW (Fig.~\ref{fig:DF}, right panel). The maximum value of the 4th Minkowski 
functional $V_3 = 20$.  In  the  $(K_1,K_2)$-plane the morphological signature 
describes a planar system \citep{e07a}.  

At a still lower density level,  $D = 4.0$, the superclusters which really do not 
belong to the SGW, join into one system with the SGW  (Fig.~\ref{fig:DF}, right 
panel).  The maximum value of the 4th Minkowski functional $V_3 = 33$, and in 
the  $(K_1,K_2)$-plane the morphological signature becomes even more typical of
a planar system, consisting of more than one individual supercluster. 
Interestingly, although the overall morphology changes strongly, 
we see that in all cases, the 
morphological signature changes at a mass fraction $m_f \approx 0.7$, a change 
of morphology at a crossover from the outskirts to the core regions \citep{e08}. 

At high mass fractions, in addition to the core of the supercluster SCl~126, also 
other high density clumps contribute to the full  SGW, thus they also add their 
contribution to the values of the 4th Minkowski functional $V_3$ and to 
$K_1$ and $K_2$ in Fig.~\ref{fig:MF}. Thus in Fig.~\ref{fig:MF} 
the values of $V_3$ for the density level, say, $D = 4.9$, at mass fractions which 
correspond to the density level $D = 6.8$ (approximately, in Fig.~\ref{fig:MF} 
this corresponds to the mass fraction $mf = 0.85$) are larger than those from the 
density level $D = 6.8$ at the same mass fraction. In other words, 
the curves characterizing the morphology of the SGW at 
different density levels depend on all the systems encompassed by the density iso-surfaces.

\subsection{Morphology of galaxy populations in the superclusters
SCl~126 and SCl~111}

Next we calculate the 4th Minkowski functional $V_3$ and the morphological 
signature $K_1$-$K_2$ for individual galaxy populations defined by  their $g - r$  
color, and for the BRGs,  in the two richest superclusters in the SGW: the 
superclusters SCl~126 (separately for the core and for the whole 
supercluster) and SCl~111 (Fig.~\ref{fig:MFG1} -- Fig.~\ref{fig:MFG3}). 
We show in figures the 95\% confidence regions obtained by Monte-Carlo simulations
as explained in sect. 3.1.

To help to understand the results of these calculations we show in 
Fig.~\ref{fig:sky68} the sky distribution of galaxies in the superclusters 
SCl~126 and SCl~111. The distribution of all galaxies in the full superclusters 
is delineated with 2D density contours. The location of groups with BRGs (we 
discuss the group membership of BRGs in the next section) is shown with 
different symbols---dots mark the members of both superclusters SCl~126
and SCl~111 at a high density 
level, $D = 6.8$, crosses mark members of another systems at this density level 
(these systems have not yet joined the superclusters SCl~126 and 
SCl~111, correspondingly), and x-s mark groups with BRGs which do not belong to 
any supercluster at the density level $D = 6.8$ yet, but will be  the members of 
the superclusters SCl~126 and SCl~111 at the density level $D = 4.9$. 

In Fig.~\ref{fig:MFG1} we present the results of the calculations of the 
Minkowsky functional $V_3$ and the shapefinders for the core of the supercluster 
SCl~126c (the groups with BRGs in this system are marked with dots in 
Fig.~\ref{fig:sky68}). Fig.~\ref{fig:MFG1} shows that the clumpiness of red and 
blue galaxies and BRGs is rather similar over the whole range of mass fractions 
$m_f$.  
The BRGs form some clumps at the mass fractions $m_f =$0.3--0.4 and $m_f =$0.5--0.7, 
which are not seen in the distribution of red and blue galaxies;
their spatial distribution is less clumpy at these densities.   
The confidence regions show that these differences are statistically significant at least
at the 95\% confidence level. 

At the mass fraction $m_f = 0.8$ 
the clumpiness of red galaxies is the largest, reaching the value $V_3 = 9$. 
The differences between the clumpiness of red galaxies and other galaxy populations
are statistically significant at least at the 95\% confidence level.
The 
clumpiness of blue galaxies and BRGs reaches the same value at the mass fraction 
$m_f = 0.80$. At a higher mass fraction, $m_f = 0.9$, blue galaxies form some 
isolated clumps and the value of $V_3$ for blue galaxies is larger than that for 
red galaxies and BRGs ($V_3 = 5$ and $V_3 \leq 3$, correspondingly).
At these highest mass fractions the differences between the clumpiness
of  blue and all other galaxy populations are statistically significant at least at the 95\%confidence level.

The morphological signatures of galaxies from different populations almost 
coincide at the mass fractions $m_f < 0.7$ (in the outskirts of the supercluster). 
The 
morphological signature for the core of the supercluster SCl~126c is 
characteristic of a filament \citep{e07a}. In this filament the distribution of 
galaxies from different populations is rather homogeneous, showing similar 
clumpiness and morphological signatures.

\begin{figure*}[ht]
\centering
\resizebox{0.95\textwidth}{!}{\includegraphics*{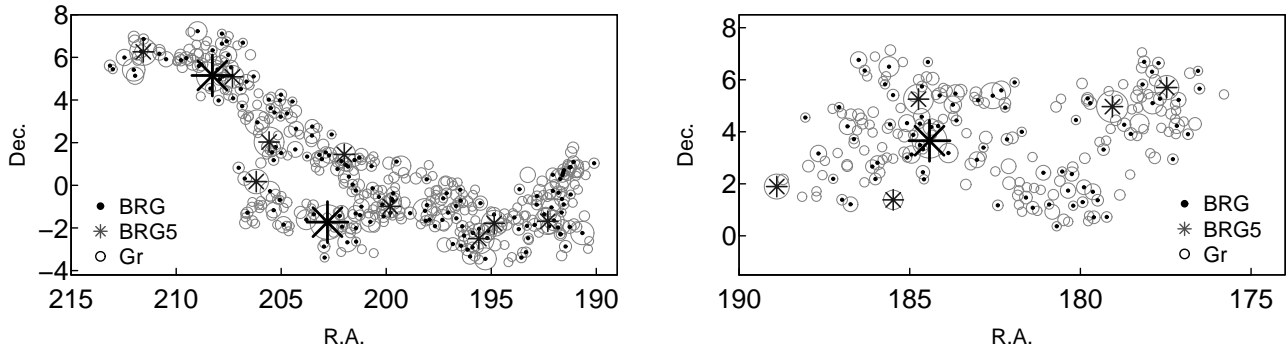}}
\caption{The distribution of groups (Gr) and BRGs 
in the superclusters SCl~126 (left panel), and
SCl~111 (right panel), in the sky. Empty circles show the location of
groups, with symbol sizes proportional to the richness of the group.
Dots mark groups which host up to four BRGs (denoted as BRG), 
stars---groups with five or more BRGs (BRG5). The three largest stars
denote groups with the largest number of BRGs in each system 
(the core of the supercluster SCL~126, SCl~126c, the outskirts of this supercluster,
SCl~126o, and the supercluster SCl~111). 
  }
\label{fig:scl6024sky}
\end{figure*}

In  Fig.~\ref{fig:MFG2} we show the 4th Minkowski functional $V_3$ and the 
morphological signature $K_1$-$K_2$ for the whole supercluster SCl~126. This 
figure reveals several differences in comparison with the core of the 
supercluster. At the mass fractions $0.15 < m_f < 0.25$ BRGs form some isolated 
clumps (for BRGs, $V_3 = 3$) while $V_3$ for red galaxies has a small minimum 
($V_3 = 0$) suggesting that at a low density level in the outskirts of the 
supercluster there are differences in how the substructure of the supercluster is 
delineated by BRGs and red galaxies. The value of the 4th Minkowski functional 
$V_3$ for red galaxies ($V_3 = 0$) suggests that the population of 
red galaxies may have tunnels through them at these mass fractions (see 
Appendix~\ref{sec:MF}). 
But the differences between the clumpiness of red galaxies and BRGs at this mass 
fraction
interval are not statistically significant at the 95\% confidence level, as show
the confidence regions in Fig.~\ref{fig:MFG2}.

At the mass fractions $m_f \approx 0.3$ the clumpiness of red galaxies and BRGs 
increases, while the clumpiness of blue galaxies decreases and even becomes 
negative. This means that at intermediate densities (up to $m_f \approx 0.6$), 
red galaxies and BRGs form isolated clumps, while blue galaxies are located in 
lower  density systems which at these mass fractions do not contribute to the 
supercluster. Some of these isolated clumps belong to a separate system, marked 
with crosses in Fig.~\ref{fig:sky68} (left panel). Negative values of  $V_3$ for 
blue galaxies at mass fractions $m_f \approx 0.45$ suggest that these systems 
have tunnels through them. These systems belong to the outskirts of the 
supercluster SCl~126. 
At this mass fraction interval the differences between the 
clumpiness of red galaxies and BRGs, and of blue galaxies,   are statistically 
significant at least at the 95\% confidence level.

At higher mass fractions, $m_f \geq 0.5$, the clumpiness of blue galaxies 
increases. The clumpiness of red and blue galaxies reaches a maximum value, 
$V_3 = 13$, at mass fractions $m_f =$0.7--0.8, where the morphological signature shows a change of the morphology of the 
supercluster. 
At still higher mass fractions we  do not see differences between 
the fourth Minkowski functionals of galaxy populations which are 
statistically significant at the 95\% confidence level.

For the most of the mass fraction interval, the clumpiness of red galaxies is larger 
than the clumpiness of blue galaxies.  The differences between the
morphological signatures of red and blue galaxies are statistically significant
for the high mass fractions at the 95\% confidence level.

Fig.~\ref{fig:MFG3} shows the 4th Minkowski functional $V_3$ and the 
morphological signature $K_1$-$K_2$ for another rich supercluster in the SGW, 
the supercluster SCl~111. The curves of $V_3$  differ  from those 
for the supercluster SCl~126. At mass fractions $0.15 < m_f < 0.35$ the $V_3$ 
curves suggest that red galaxy and BRG populations have tunnels through them. 
At these mass fractions some 
galaxies already do not contribute to the supercluster (these are galaxies which 
do not belong to any supercluster at higher density levels in 
Fig.~\ref{fig:sky68}) so that tunnels can form. In the same time individual 
clumps are connected, so that tunnels form between them (otherwise red galaxies 
and BRGs would form isolated clumps instead of tunnels). At mass fractions  $m_f 
> 0.35$ the  value of  $V_3$ for red galaxies and BRGs increases showing that 
the number of isolated clumps formed by these galaxies grows. The $V_3$ 
curves for red galaxies and BRGs reach a maximum value ($V_3 =$9--10) at a mass 
fraction $m_f \approx 0.6$ and then decrease. Such a clumpiness curve is 
characteristic of a poor supercluster, a system dominated by a few high density 
clumps with rich clusters in it \citep[a spider, see ][]{e07a}. 
Fig.~\ref{fig:sky68} (right panel) shows low density regions (groups which do 
not belong to any supercluster at densities $D = 6.8$) and  high density regions 
which form systems already at densities $D = 6.8$ in this supercluster. 

The value of the  4th Minkowski functional $V_3$ for blue galaxies in the 
supercluster SCl~111 $V_3 = 1$ up to mass fractions  $m_f < 0.3$ shows that 
at these mass fractions blue galaxies form one system without isolated clumps or 
tunnels through them. In other words, in the outskirts of this supercluster blue 
galaxies are distributed more homogeneously than red galaxies or BRGs. In the 
mass fraction interval $0.35 < m_f < 0.45$ the value of $V_3$ for blue galaxies 
decreases---some blue galaxies do not contribute to supercluster any more and 
tunnels form.  At the mass fraction $m_f > 0.45$ the value of $V_3$ increases (the 
number of isolated clumps increases) and reaches the maximum ($V_3 = 9$) at the 
mass fraction $m_f = 0.8$. After that, the clumpiness of blue galaxies 
decreases. 

In Fig.~\ref{fig:MFG3} we see that at almost the whole mass fraction
interval the differences between the values of the 4th Minkowski functional $V_3$
for galaxies from different populations are statistically significant 
at least at the 95\% confidence level.

The morphological signature of the supercluster SCl~111 also differs from that 
of the supercluster SCl~126. The values of the planarity $K_1$ at small mass 
fractions are larger than those for the SCl~126, and the maximum values of the 
filamentarity $K_2$ are lower ($K_2 \leq 0.7$, while for the supercluster 
SCl~126 the filamentarity $K_2 \leq 0.9$). Interestingly, for certain mass 
fractions ($0.45 < m_f < 0.65$) the planarity of BRGs is larger than that of red 
and blue galaxies. This may be due to the large clumpiness of BRGs in the same 
mass fraction interval (Fig.~\ref{fig:MFG3}, left panel).

Fig.~\ref{fig:MFG3} shows that  the differences between 
morphological signatures of red galaxies, BRGs
 and blue galaxies are statistically significant
at the 95\% confidence level at a wide range of mass fractions.

\subsection{Summary}

We have seen that the morphology of the SGW varies from  a simple filament to a 
multibranching filament and then,  as we lower the level of the 
density, becomes characteristic of a very clumpy, planar 
system. The changes in morphology suggest, in addition to the analysis of the 
properties of individual superclusters (their richnesses and  sizes), that a 
proper density level to delineate individual superclusters is $D = 4.9$. A lower 
density level, $D = 4.7$, can be used to select the whole SGW. 

The overall morphology of the galaxy systems in the SGW region changes strongly 
when we change the density level, but in all 
cases, the morphological signature changes at a mass fraction $m_f \approx 0.7$, 
at a crossover from the outskirts to the core regions of 
superclusters. 

The morphology of  the core of the supercluster SCl~126 is characteristic of a 
simple filament with small clumpiness. Differences between the clumpiness and 
morphological signatures of individual galaxy populations are small, telling us 
that  all galaxy populations delineate the core of the supercluster in the same 
way. The morphology of the whole supercluster SCl~126 can be described as 
characteristic of a multibranching filament, where the clumpiness of red 
galaxies and BRGs is larger than the clumpiness of blue galaxies, especially in 
the outskirts of the supercluster. The supercluster SCl~111 resembles a 
multispider  \citep[we described the morphological templates in more detail 
in][]{e07a}. Here again the clumpiness of red galaxies and BRGs is larger than 
the clumpiness of blue galaxies. In the outskirts of both superclusters blue 
galaxies are distributed more homogeneously than red galaxies and BRGs. In
the outskirts of superclusters galaxy populations have tunnels through them.

The morphology of superclusters can be characterized by the ratio of the
shapefinders, $K_1/K_2$ (the shape parameter). This approach was used, for example, 
by \citet{bas03} to study the shapes of superclusters--- a large value of this ratio 
indicates planar systems. Our calculations show 
that the values of  the shape  parameter are as follows: 
for the core of the supercluster SCl~126, SCl~126c, $K_1/K_2 \approx 0.33$ and for 
the full supercluster SCl~126 $K_1/K_2 \approx 0.28$; we get the same value 
for all galaxy populations. In the supercluster SCl~111, 
$K_1/K_2 = 0.47$ for the full supercluster and for red galaxies, while  for BRGs 
$K_1/K_2 = 0.52$ and for blue galaxies $K_1/K_2 = 0.61$. This is another 
difference between the superclusters in the SGW.

We assessed the statistical significance of the differences between
the 4th Minkowski functional $V_3$ and morphological signatures
of different galaxy populations in superclusters using Monte-Carlo simulations
and showed that at least at some mass fraction intervals (depending
on the populations) the differences between the galaxy populations
are statistically significant at
least at the 95\% confidence level.

\section{BRGs and the first ranked galaxies in the superclusters SCl~126 and SCl~111}
\label{sect:scl126111}

{\scriptsize
\begin{table*}[ht]
\centering
\caption{The properties of groups 
with and without BRGs in the SCl~126 and SCl~111.
}
\begin{tabular}{lrrrr|rrrr} 
\hline 
(1)&\multicolumn{4}{c}{(2)}&\multicolumn{4}{c}{(3)} \\      
\hline 
Region       &  \multicolumn{4}{c}{Gr, BRG}    &  \multicolumn{4}{c}{Gr, --}  \\     
  SCl ID     & $N_{\mbox{gr}}$ & $N_{\mbox{gal, m}}$ & $\sigma_{\mbox{m}}$& $L_{\mbox{tot, m}}$  &$N_{\mbox{gr}}$   &$N_{\mbox{gal, m}}$ & $\sigma_{\mbox{m}}$& $L_{\mbox{tot, m}}$ \\
\hline
               &  \multispan4     &  \multispan4   \\     
SCl~126c       &   67     &  12.1       &  250.5      &  20.2        & 98         & 2.9            &  135.8      &  3.5        \\      
SCl~126o       &  101     &   9.1       &  209.0      &  13.7        & 112        & 2.9            &  125.9      &  3.6        \\ 
SCl~111        &   85     &   9.6       &  199.2      &  13.1        &  116       & 2.6           &  138.2      &   2.7        \\ 
       \\
\label{tab:GRBRG}                        
\end{tabular}
\tablecomments{
The columns in the Table are as follows:
\noindent Column 1: Region and supercluster ID (as defined in the text, Sect. 2.1).
Column 2: Groups with BRGs, 
column 3: Groups without BRGs: 
$N_{\mbox{gr}}$---number of groups,
$N_{\mbox{gal, m}}$---mean richness, 
$\sigma_{\mbox{m}}$---mean rms velocity  (km/s),
$L_{\mbox{tot}}$---mean total luminosity ($10^{10} h^{-2} L_{\sun}$).
}
\end{table*}
}

\begin{figure}[ht]
\centering
\resizebox{0.47\textwidth}{!}{\includegraphics*{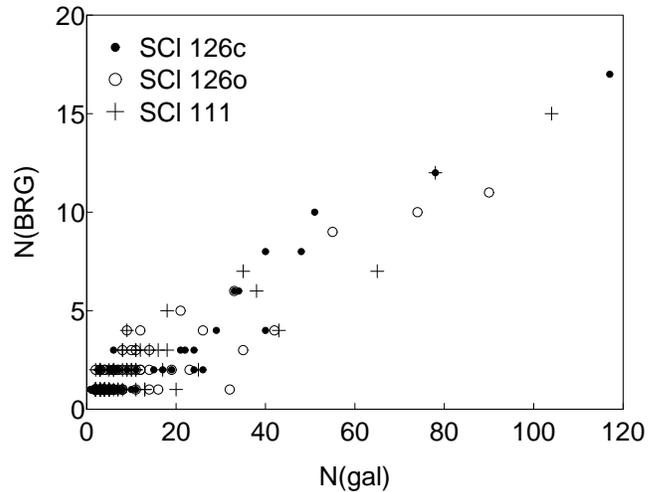}}
\caption{The number of BRGs in groups, $N_{BRG}$, in the 
superclusters SCl~126 (separately for the core, SCl~126c 
and for the outskirts, SCl~126o) and SCl~111 
vs the richness of groups (the number of 
galaxies in groups, $N_{gal}$), where the BRGs are located.
Filled circles---SCl~126c, empty circles---SCl~126o, crosses---SCl~111.
  }
\label{fig:scl126111nbrggr}
\end{figure}

As the Minkowski functionals found in the previous section demonstrated, 
the morphology of the superclusters SCl~126 and SCl~111 is different. The 
supercluster SCl~126, especially its high density core region, is a 
multibranching filament of a high overall density. The supercluster SCl~111 
consists of three loosely located, less dense concentrations (see also 
Fig.~\ref{fig:sky68}).  The Minkowski functionals show several 
similarities and differences between the way how the substructure of superclusters 
is delineated by different galaxy populations. 

In this section  we focus on the comparison of galaxy and group content in different 
superclusters of the SGW to search for other possible differences between them, in 
addition to the differences in morphology. We study the properties of BRGs and 
those first ranked galaxies, which 
are not BRGs. In the case of the 
supercluster SCl~126 we consider separately the core region, SCl~126c, and the
outskirts, SCl~126o. We analyze the group membership of BRGs, as well as the 
colors, luminosities and morphological types of BRGs in the superclusters SCl~126 
and SCl~111, calculate the peculiar velocities of BRGs and other first ranked 
galaxies in groups, and study the location of groups with different first ranked 
galaxies in superclusters.

\subsection{Group membership of BRGs}

Fig.~\ref{fig:scl6024sky} shows the sky distribution of the BRGs and galaxy 
groups in the superclusters SCl~126 (left panel) and SCl~111 (right panel). Here 
we plotted groups with symbols  of size proportional to their richness. 
In this figure we mark those groups which host up to four BRGs with 
small dots and groups with at least five BRGs---with stars. The comparison with 
Fig.~\ref{fig:sky68} shows that in the supercluster SCl~126 groups with a large 
number of BRGs  are located both in the core and in the outskirts. 
In the supercluster SCl~111 assemblies of groups with a large
number of BRGs form systems already at the density level $D = 6.8$, while
the poor groups and groups with a few  BRGs in them
do not belong to any supercluster at this density level.

{\scriptsize
\begin{table*}[ht]
\centering
\caption{BRGs and the first ranked galaxies in the SCl~126 and SCl~111.
}
\begin{tabular}{lrrr|rrr|rrr} 
\hline 
(1)&\multicolumn{3}{c}{(2)}& \multicolumn{3}{c}{(3)} &  \multicolumn{3}{c}{(4)}\\      
\hline 
Population ID&\multicolumn{3}{c}{SCl~126c} &\multicolumn{3}{c}{SCl~126o}   &  \multicolumn{3}{c}{SCl~111}     \\ 
\hline    
$N_{\mbox{gal}}$    &      & 1397 &          &      & 1765 &      &      & 1515 &          \\
$f_{\mbox{r}}$     &      & 0.72 &          &      & 0.61 &      &      & 0.63 &          \\
$ID           $    &   B1 &  Bs  &  1nB     & B1   &  Bs  &  1nB &  B1  &  Bs  &  1nB     \\
$N_{\mbox{gal}}$    &   67 &  114 &   98     & 101  & 140  & 159  &  85  &  123 &  135     \\
$f_{\mbox{r}}$     & 0.93 & 0.96 &  0.64    & 0.93 & 0.87 & 0.49 & 0.93 & 0.93 &  0.47 \\
$f_{\mbox{s}}$     & 0.39 & 0.42 &  0.59    & 0.35 & 0.58 & 0.82 & 0.45 & 0.51 &  0.77 \\ 
$f_{\mbox{rs}}$    & 0.34 & 0.39 &  0.44    & 0.31 & 0.52 & 0.71 & 0.41 & 0.48 &  0.57 \\
$N_{\mbox{merg}}$   &    9 &   14 &     1    &    3 &   10 &    4 &   13 &   12 &     6 \\
$N_{\mbox{clock}}$  &    6 &    4 &     9    &    8 &    5 &   21 &    7 &    3 &    16 \\ 
$N_{\mbox{a-clock}}$&    4 &   15 &    14    &    9 &    6 &   21 &    4 &   13 &    16 \\
$W_{\mbox{E}}$     & 0.031& 0.030&  0.036   & 0.029& 0.027& 0.066& 0.031& 0.034&  0.073\\ 
$W_{\mbox{S}}$     & 0.060& 0.052&  0.035   & 0.047& 0.060& 0.039& 0.042& 0.058&  0.058\\
$vE_{\mbox{pec}}$   & 186  & 467  &  132     & 141 & 260  & 133  & 154  & 269  &  178  \\
$vS_{\mbox{pec}}$   & 154  & 285  &  109     & 205 & 290  & 101  & 164  & 397  &  105  \\
       \\
\label{tab:sclgal}                        
\end{tabular}
\tablecomments{
The columns in the Table are as follows:
\noindent Column 1: Population ID.
$f_{\mbox{r}}$: fraction of red galaxies ($g - r > 0.7$),
$f_{\mbox{s}}$: fraction of spiral galaxies,
$f_{\mbox{rs}}$: fraction of spirals among red galaxies,
$N_{\mbox{merg}}$: number of galaxies, which show signs of merging or other disturbancies,
$N_{\mbox{clock}}$: number of spirals with clockwise spiral arms,
$N_{\mbox{a-clock}}$: number of spirals with anti-clockwise spiral arms,
$W_{\mbox{E}}$: $g-r$ rms color scatter for elliptical BRGs,
$W_{\mbox{S}}$: rms color scatter for spiral BRGs,
$vE_{\mbox{pec}}$ and $vS_{\mbox{pec}}$: the mean peculiar velocities
of elliptical and spiral galaxies for a given population, in km/s. 
Columns 2--4:  supercluster  ID and population ID, where B1 means first ranked 
BRGs, Bs---those BRGs which are not first ranked (satellite BRGs),
and 1nB indicates those first ranked galaxies, which do not belong to BRGs.
}
\end{table*}
}

The richest group in the core of the supercluster SCl~126, that corresponds to 
the Abell cluster A1750, located at $R.A \approx 202.7^\circ$ and $Dec \approx -1.9^\circ$ 
(Fig.~\ref{fig:scl6024sky}, left panel) hosts a full 17 BRGs. In the outskirts of 
the supercluster SCl~126 the number of BRGs is the largest in a group that 
corresponds to the Abell cluster A1809, located at $R.A \approx 208.3^\circ$ and $Dec 
\approx 5.1^\circ$, this group hosts 12 BRGs; isolated clumps seen in the $V_3$ curve 
at mass fractions of about $m_f \approx 0.2$ (Fig.~\ref{fig:MFG2}) are probably 
at least partly due to this group. In the supercluster SCl~111, the richest 
group corresponds to the Abell cluster A1516, located at $R.A \approx 184.4^\circ$ and 
$Dec \approx 3.7^\circ$, and hosts 15 BRGs. The location of the groups with the 
largest numbers of BRGs in Fig.~\ref{fig:scl6024sky} is marked with the largest stars. 

In Fig.~\ref{fig:scl126111nbrggr} we compare the number of BRGs
in groups in the superclusters SCl~126 (in the core and in the outskirts)
and in the supercluster SCL~111. This figure shows that most groups
with BRGs host only a few BRGs (the mean number of BRGs in groups is 2
in both the superclusters SCl~126 and SCl~111). Groups with  five or more 
BRGs are themselves also richer than groups with a smaller
number of BRGs, with at least 20  member galaxies.
Although the richest group in the core of the supercluster SCl~126c
hosts more BRGs than the richest group in the outskirts of this supercluster,
the overall membership of BRGs in groups is statistically similar in
both these regions, and also in the supercluster SCl~111.
Table~\ref{tab:GRBRG} shows that approximately 40\% of BRGs are first ranked
galaxies in groups. 

There are also BRGs, which do not belong to any group (isolated BRGs).
In the core of the supercluster SCl~126, SCl~126c, 11\% of all BRGs 
do not belong to groups, while in the outskirts of this supercluster 
this fraction is 18\%.
In the supercluster SCl~111 20\% of all BRGs do not belong to groups. 
In this respect there is a clear difference between superclusters. 

\citet{tempel09} showed, analysing the luminosity 
functions of galaxies in the 2dFGRS groups, that most of the bright isolated 
galaxies should be the first ranked galaxies of groups, where the fainter  members  
are not observed since they fall outside of the observational 
window of the flux-limited sample. Thus we may suppose that there are no truly isolated BRGs,
but they are the first ranked galaxies of fainter groups.

Fig.~\ref{fig:scl6024sky} shows that there are also many groups without BRGs (empty 
circles without a dot or a star). In the supercluster SCl~111 many of these groups 
are located loosely between the three concentrations of this supercluster 
(Fig.~\ref{fig:scl6024sky}, right panel). We present the summary of the 
properties of groups with and without BRGs in superclusters in 
Table~\ref{tab:GRBRG}. This table shows, firstly, that groups with BRGs are 
richer, have larger velocity dispersions and higher luminosities than groups 
without BRGs. Secondly,  groups with BRGs in the core of the supercluster 
SCL~126, SCl~126c, are richer and have larger velocity dispersions and higher 
luminosities than groups with BRGs 
in the outskirts of this supercluster or groups in the 
supercluster SCl~111. The Kolmogorov-Smirnov test shows that the probability 
that the distributions of the properties   of groups with BRGs in 
SCl~126c and  SCl~126o are drawn from the same distribution is 0.04,  and 0.01 
for groups in SCl~126c and SCl~111.  This shows another difference between these 
systems.

\subsection{Colors, luminosities and morphological types of BRGs and the first ranked
galaxies}

\begin{figure*}[ht]
\centering
{\resizebox{0.22\textwidth}{!}{\includegraphics*{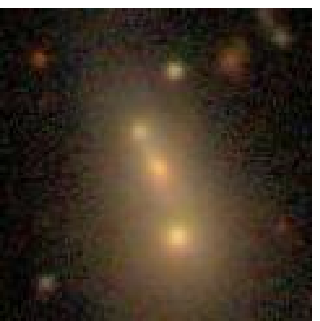}}}
{\resizebox{0.22\textwidth}{!}{\includegraphics*{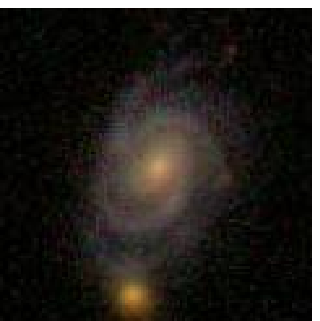}}}
{\resizebox{0.22\textwidth}{!}{\includegraphics*{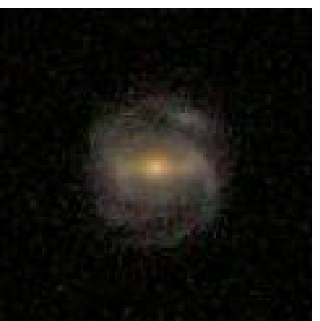}}}
{\resizebox{0.22\textwidth}{!}{\includegraphics*{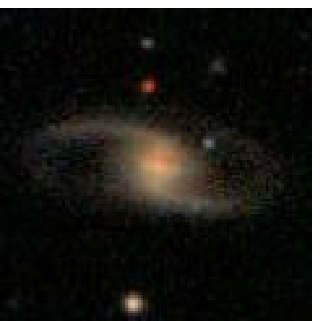}}}
\hspace*{2mm}\\
\caption{
SDSS images of some of the first ranked BRGs. From left to right: two galaxies from
the core of the supercluster SCl~126
(with ID J131236.98-011151 and J133111.02-014338.2), a galaxy from the outskirts
of the supercluster SCl~126 (J135143.99+045833.6), 
and a galaxy from the supercluster SCl~111 (J120652.46+035952.8).
}
\label{fig:images}
\end{figure*}

\begin{figure}[ht]
\centering
{\resizebox{0.48\textwidth}{!}{\includegraphics*{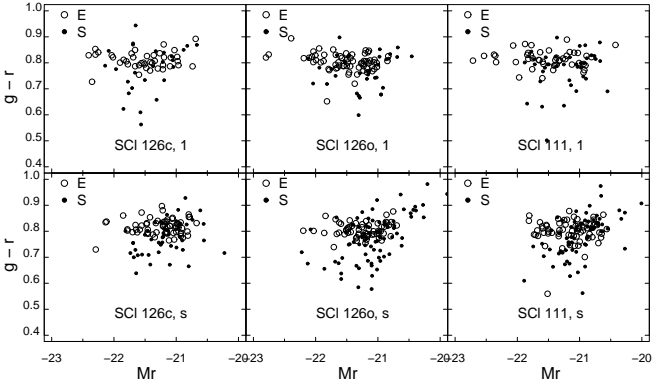}}}
\hspace*{2mm}\\
\caption{
The color-magnitude relations for 
elliptical (empty circles) and spiral (filled circles), 
first ranked (upper row) and satellite (lower row)
BRGs.  Left panels: the core of the supercluster SCl126,
middle panels: the outskirts of the supercluster SCl126, right panels:
the supercluster SCl111.
}
\label{fig:cmr6}
\end{figure}

Next we divide BRGs into two populations: the first ranked galaxies in 
groups and those which are not
(we denote them as satellite BRGs). We compare the 
colors, luminosities and morphological types of these two
sets of BRGs and those first 
ranked galaxies which are not BRGs in the two richest superclusters in 
the SGW, SCl~126 and SCl~111.

We summarize the colors and morphology of the galaxies of the two 
superclusters in Table~\ref{tab:sclgal}. It shows that most BRGs are 
red, although a small fraction of BRGs have blue colors. The fraction 
of red galaxies among these first ranked galaxies, which are not BRGs, 
is smaller. We note that in the core of the supercluster SCl~126
the fraction of red galaxies among those first ranked galaxies which
are not BRGs is larger and the fraction of spirals 
(and red spirals) among them is smaller
than in the outskirts of this supercluster
or in the supercluster SCl~111.

A large fraction of red galaxies in our sample are spirals. The images of 
some BRGs in the SGW are shown in Fig.~\ref{fig:images}. Galaxies with visible 
spiral arms are located in groups of different richness, and more than a half of 
them do not belong to any group. We see also that a considerable number of 
BRGs and the first ranked galaxies may be merging or have traces of past merging 
events.

The scatter plots of the colors of ellipticals and spirals 
among the first ranked and satellite BRGs are 
shown in Fig.~\ref{fig:cmr6}  (see also Table~\ref{tab:sclgal}).  
Both  show that for 
the red elliptical first ranked galaxies the color scatter is small 
(about 0.03), while for the red spirals it is larger (0.04--0.06). 
Some satellite BRGs have blue colors.
Fig.~\ref{fig:cmr6} shows that most BRGs with blue colors are spirals, but
there are also two elliptical galaxies among them.
  
\begin{figure}[ht]
\centering
{\resizebox{0.48\textwidth}{!}{\includegraphics*{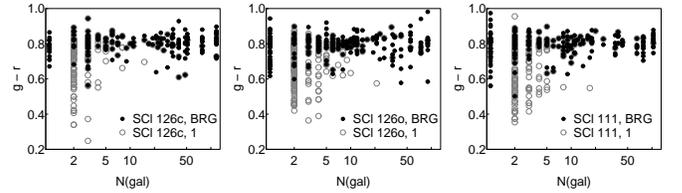}}}
\hspace*{2mm}\\
\caption{
The $g - r$  
color  vs the group richness of BRGs (filled circles) 
and the first ranked galaxies which are not BRGs (empty circles).
Left panel: the core of the supercluster SCl126,
middle panel: the outskirts of the supercluster SCl126, right panel:
the supercluster SCl111.
}
\label{fig:BRG1stnggr}
\end{figure}

\begin{figure}[ht]
\centering
{\resizebox{0.48\textwidth}{!}{\includegraphics*{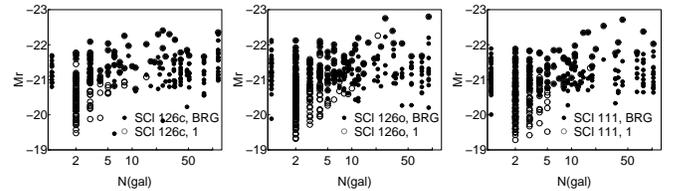}}}
\hspace*{2mm}\\
\caption{
The  luminosity vs the group richness of BRGs (filled circles) 
and the first ranked galaxies which are not BRGs (empty circles).
Left panel: the core of the supercluster SCl126,
middle panel: the outskirts of the supercluster SCl126, right panel:
the supercluster SCl111.
}
\label{fig:BRG1stngmr}
\end{figure}

Fig.~\ref{fig:BRG1stnggr} and Fig.~\ref{fig:BRG1stngmr}
show the colors and  luminosities
of BRGs and those first ranked galaxies, which are not BRGs, vs the 
number of galaxies in the groups where they reside. 
We see that the first ranked galaxies with bluer colors and also 
fainter first ranked galaxies and fainter BRGs 
reside in poor groups. In almost all groups with $N_{gal} > 10$
the first ranked galaxy is a BRG.
The Kolmogorov-Smirnov test shows that
the probabilities that the colors and luminosities of BRGs and the
first ranked galaxies in systems under study are drawn from the same
distributions are very small.

Comparison of the richnesses of groups where elliptical or spiral BRGs
reside shows that almost all the spiral first ranked BGRs belong to poor groups
with less than 13 member galaxies. At the same time spiral, satellite BRGs can be
found in groups of any richness.

Our results are consistent with those by \citet{berlind} who showed that the
first ranked galaxies in more luminous groups are more luminous than the first ranked
galaxies in less luminous groups
\citep[see also][and references therein]{2008ApJ...676..248Y,tempel09}.

\subsection{Peculiar velocities of BRGs and the first ranked galaxies}

Next we calculated  the peculiar velocities for the red ($g-r>0.7$) elliptical and 
spiral BRGs in three systems under study, for three sets of galaxies (the first 
ranked BRGs, satellite BRGs, and those first ranked galaxies, which are not 
BRGs) (Table~\ref{tab:sclgal}). 

Our calculations show that in both the superclusters SCl~126 and SCl~111 about half 
of the first ranked galaxies have large peculiar velocities. The largest 
peculiar velocities of the first ranked galaxies are even of the order of 1000 
km/s. 

In the supercluster SCl~126c, the peculiar velocities of elliptical BRGs are 
larger than those of spiral galaxies (both in the case of the first ranked galaxies 
and satellite BRGs). The statistical significance of differences of peculiar 
velocities between the elliptical and spiral first ranked galaxies is low, according 
to the Kolmogorov-Smirnov test. The differences between the peculiar velocities 
of elliptical and spiral satellite galaxies are statistically significant at the 
95\% confidence level. 

In the superclusters SCl~126o and SCl~111  elliptical 
BRGs have smaller peculiar velocities than spiral BRGs, but those
elliptical first ranked galaxies which do not belong to BRGs have
larger peculiar velocities than spiral galaxies. The 
peculiar velocities of BRGs in these two superclusters are smaller 
than in the supercluster SCl~126c. According to the Kolmogorov- Smirnov test, these 
differences are statistically significant at the 92\% level in both 
superclusters.

\begin{figure*}[ht]
\centering
\resizebox{0.95\textwidth}{!}{\includegraphics*{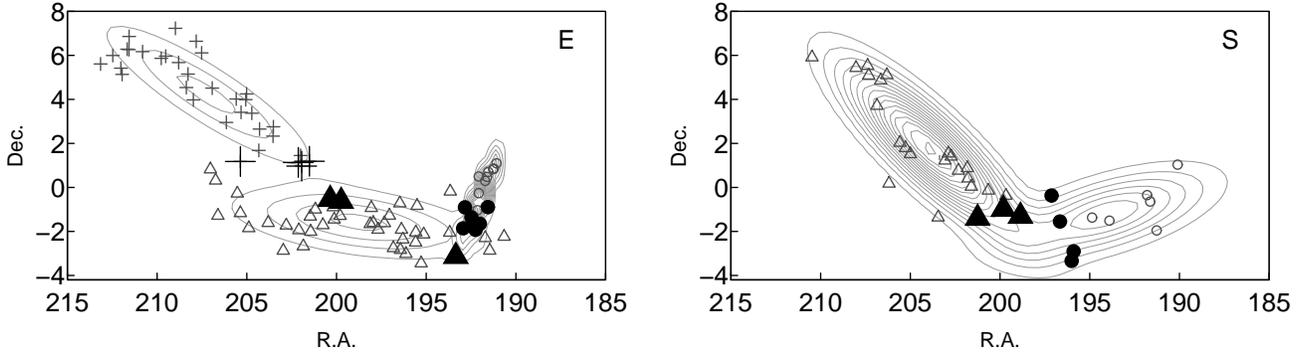}}
\caption{The sky distribution of groups with at least 3 member galaxies 
with BRGs as their first ranked 
galaxy, in the supercluster SCl~126.
Left panel---groups with  
elliptical first rank BRGs, right panel---groups with spiral first ranked BRGs.
Different symbols correspond to different components determined with 'Mclust',
large symbols (in the case of 
triangles and circles - large filled symbols) correspond to these groups for 
which the uncertainty of the classification is larger than 0.05.
Filled circles show the locations of groups, contours show the projected density levels
calculated with 'Mclust' (see text).
  }
\label{fig:scl60essky}
\end{figure*}

\begin{figure*}[ht]
\centering
\resizebox{0.95\textwidth}{!}{\includegraphics*{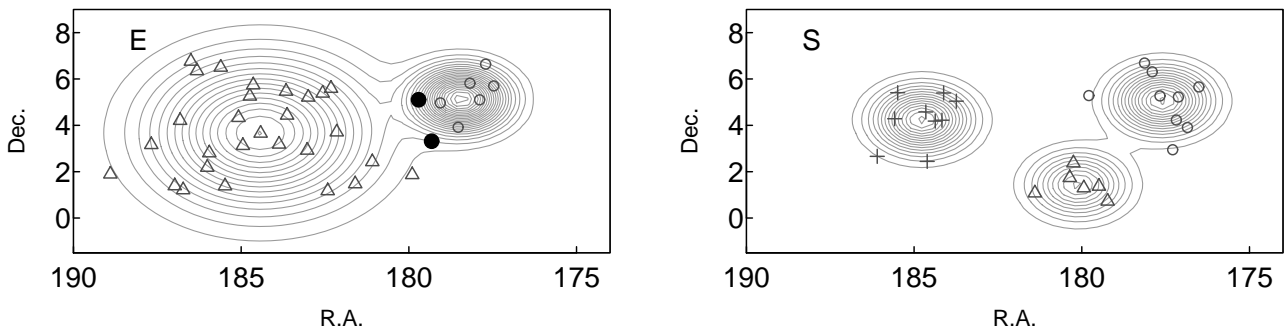}}
\caption{The sky distribution of groups with at least 3 member galaxies
with BRGs as their first ranked 
galaxy, in the supercluster SCl~111.
Left panel---groups with  
elliptical first rank BRGs, right panel---groups with spiral first ranked BRGs.
Symbols are as in Fig.~\ref{fig:scl60essky}. 
Contours show the projected density levels.
  }
\label{fig:scl24essky}
\end{figure*}

\subsection{Sky distribution of groups with elliptical
and spiral BRGs as the first ranked galaxy}

Now we compare the distribution of groups with elliptical and spiral BRGs as 
their first ranked galaxies in the sky
(Fig.~\ref{fig:scl60essky} and Fig.~\ref{fig:scl24essky}). 
These figures show that in both superclusters the groups with 
elliptical first ranked BRGs fill the supercluster volume in a much more uniform 
manner than the groups with the spiral first ranked BRG.
This was unexpected, since 
groups with an elliptical BRG as the first ranked galaxy are richer, and,
as we showed, groups in the core of the supercluster SCl~126 are richer than in
 other systems under study. 


To quantify the distribution  of groups in 
superclusters (substructure of superclusters)
and to calculate the projected density contours, we have used the R package 
'Mclust' \citep{fraley2006}. 
This package builds normal mixture models for the data, 
using multidimensional Gaussian densities with varying shape, 
orientation and volume. It finds the optimal number of concentrations,
the corresponding classification 
(membership of each concentration), and the uncertainty of the classification.
The uncertainty of the classification is  defined using the probabilities 
for each galaxy to belong to a particular component and  
is calculated as 1. minus the highest probability of a galaxy to belong to a component.
The uncertainty of the classification can be used
as a statistical estimate  
how well groups are assigned to the components. 

To measure how well components are determined and to find the best model for a 
given dataset, 'Mclust' uses the Bayes Information Criterion (BIC) that is based 
on the maximized loglikelihood for the model, the number of variables and the number of 
mixture components. The model with the smallest value of BIC among all models 
calculated by 'Mclust' is considered the best. For details we refer to
\citet{fraley2006}.

We present the results in Fig.~\ref{fig:scl60essky} and 
Fig.~\ref{fig:scl24essky}. In these figures we plot components of the best model 
according to BIC, determined by 
'Mclust' in the distribution of groups with elliptical and spiral BRGs as the 
first ranked galaxies with different symbols, large symbols (in the case of 
triangles and circles - large filled symbols) correspond to these groups for 
which the uncertainty of the classification is larger than 0.05.
Groups with spiral BRGs as the 
first ranked galaxies in the supercluster SCl~111 all have the 
uncertainty of the classification smaller than 0.05. 

Fig.~\ref{fig:scl60essky} and 
Fig.~\ref{fig:scl24essky} show that 
groups with elliptical and spiral BRGs as the first ranked galaxies populate 
superclusters in a different manner. In the supercluster SCl~126 groups with 
elliptical BRGs as the first ranked galaxies form three concentrations in the 
supercluster, and groups with a spiral BRG as the first ranked galaxy  form two 
concentrations. The mean uncertainty of the classification calculated by
'Mclust' for the groups with 
elliptical first ranked BRGs is $2.0\cdot10^{-3}$, 
for the groups with spiral first ranked BRGs
it is $2.0\cdot10^{-4}$.  
Very small uncertainties show very
high probabilities for groups to belong to a particular component.
Groups with the largest uncertainties of the classification are located 
there where components are overlapping.

For every dataset we analyzed in detail three best models found by 'Mclust'.  
For groups with elliptical BRGs as the first ranked galaxies in the supercluster
SCl~126 two best models have three, the third even four components. The best 
model with two components for these groups is, according to BIC, very unlikely,
and in this case 'Mclust' collects together into one component
groups indicated with circles and triangles in Fig.~\ref{fig:scl60essky}.
These components differ from those determined using 
groups with spiral BRGs as the first ranked galaxies in this supercluster
Fig.~\ref{fig:scl60essky}.

In the supercluster SCl~111 the groups with  an elliptical BRG as the first ranked galaxy 
form two barely separated concentrations in the supercluster, 
and with spiral first ranked BRGs---three well-defined 
concentrations (Fig.~\ref{fig:scl24essky}). The mean uncertainty of the 
'Mclust' classification for the groups with elliptical first ranked BRGs 
is $9.4\cdot10^{-9}$, for the groups with spiral first ranked BRGs $7.5\cdot10^{-7}$.

The best model with three components for groups with elliptical BRGs 
as the first ranked galaxies in the supercluster SCl~111 is, according to BIC,
much less likely.
In this model groups from the component 
shown by triangles in Fig.~\ref{fig:scl24essky}
are divided between two components, the leftmost groups forming a separate 
component. Thus components differ from those
determined in the distribution of groups with spiral BRGs as the 
first ranked galaxies in this supercluster.


If the components have a regular shape (ellipsoids or circles)  as in the 
supercluster SCl~111 (Fig.~\ref{fig:scl24essky}, the component with mean 
coordinates $R.A \approx 184.4^\circ$ and $Dec \approx 3.67^\circ$) the scatter of the sky 
coordinates can be used to estimate the compactness/looseness of the sky 
distribution of groups in components. The rms scatter of coordinates in the case of 
groups with elliptical first ranked BRGs is $\sigma \approx 3.9^\circ$, for groups 
with spiral first ranked BRGs---$\sigma \approx 0.8^\circ$ that shows a more compact 
distribution of groups with spiral first ranked galaxies in this component. 
However, in the supercluster SCl~126 the sky distribution of groups is such that 
the scatter of coordinates of all components is large 
(Fig.~\ref{fig:scl60essky}) so we cannot use this estimate. In this supercluster 
we compared the fraction of groups with spiral and elliptical first ranked BRGs 
in the core and the outskirts of the supercluster. A full 47\% of groups with a spiral first ranked BRG lie in the core region, while among groups with an elliptical BRG 
as the first ranked galaxy only 39\% lie in the core region. 

The number of groups with an elliptical first ranked galaxy in both superclusters
is larger than the number of groups with a spiral first ranked galaxy.
Thus we can resample groups with an elliptical first ranked galaxy by
diluting randomly the sample of groups so that in the final sample
the number of groups is equal to the number of groups with a spiral first 
ranked galaxy, and then run 'Mclust' again to search for components 
among the resampled set of groups. However, a word of caution is needed -- if
'Mclust' will detect a different number of components
among the randomly resampled groups compared to the original sample, 
the sky coordinates of these components depend on this particular realization
of the resampling. We run the resampling procedure for both
superclusters 10000 times. Our calculations showed that 
in the supercluster SCl~126, in about 27\%
of all cases 'Mclust' detected in the distribution of groups  
with an elliptical first ranked galaxy two components, as in the case of 
the groups with a spiral first ranked galaxy, but the sky coordinates 
of these components varied strongly, covering the whole right acsension
and declination interval of the supercluster SCl~126. The same is the
case with the supercluster SCl~111, where 'Mclust' detected 
in 14\% of cases three components in resampled groups with an
elliptical first ranked galaxy. 
This shows that when groups with an elliptical first ranked galaxy
are resampled to the same number of objects as groups with spiral
first ranked galaxy there is no clear difference in number of components
between samples. 

To understand better the differences between the distribution of groups with 
elliptical and spiral first ranked galaxies in superclusters a study of a larger 
sample of superclusters is needed.


We note that \citet{berlind} showed that the groups with first ranked galaxies 
with bluer colors are more strongly clustered than the groups with the first 
ranked galaxies with redder colors. In contrast, \citet{2008ApJ...687..919W} 
showed that SDSS groups with red central galaxies  are more strongly clustered 
than groups of the same mass but with blue central galaxies. As we showed, 
colors and morphological types of galaxies are not well correlated. Therefore, 
we plan to study a larger sample of groups with the with the first ranked 
galaxies of different morphological type and color in a larger set of 
superclusters to search for possible differences in their distribution. 

\subsection{Summary}

The  comparison of  the group membership and properties (colors, luminosities and
morphological types), as well as the peculiar velocities and the sky
distribution of BRGs and those first ranked galaxies which are
not BRGs in the two SGW superclusters revealed a number of
differences between these superclusters.

In all three systems under study (the core and the outskirts of the supercluster
SCl~126, and the supercluster SCl~111) rich groups with at least 20 member
galaxies host at least five BRGs, the richest group in the
supercluster SCl~126 in the core of the supercluster hosts even 17 BRGs.
The fraction of those BRGs which do not belong to any group 
in the core of the supercluster SCl~126 (SCl~126c), is about two
times smaller than in other systems under study.  
 
In the core of the supercluster
SCl~126 (SCl~126c), groups with BRGs are richer and have larger velocity 
dispersions than groups with BRGs in the outskirts of this supercluster
and in the supercluster SCl~111.

In the core of the supercluster
SCl~126 (SCl~126c) the fraction of red galaxies among those first
ranked galaxies, which are not BRGs, is larger 
and the fraction of spirals 
among them is smaller than in the outskirts
of this supercluster and in the supercluster SCl~111.

About 1/3 to 1/2 of all BRGs are spirals. The scatter of the $g - r$ colors of 
elliptical BRGs is about two times smaller than of spiral BRGs, showing that 
elliptical BRGs form a much more homogeneous population than spiral 
BRGs. 

Peculiar velocities of elliptical BRGs in the core of the supercluster SCl~126 
(SCl~126c) are larger than those of spiral BRGs, while in the outskirts of this 
supercluster, as well as in the supercluster SCl~111, the situation is opposite. 
Since both elliptical and spiral satellite BRGs lie in groups of all richnesses, 
this difference is not caused by the richness of the host group (this could be 
the case if spiral satellites were residing preferentially in poor groups). In 
the core of the supercluster SCl~126 (SCl~126c) the peculiar velocities of 
galaxies are larger than in the superclusters SCl~126o and SCl~111 suggesting 
that groups in the core of the supercluster SCl~126 are dynamically more active 
than those in other two systems under study. 

In the two superclusters, groups with an elliptical first ranked BRG 
are located more uniformly than groups with a spiral first ranked BRG.

We note that, as we saw in section 3.3, 
although the outskirts of the supercluster SCl~126o and
the supercluster SCl~111 are determined by a lower
density level than the core of the supercluster SCl~126c, they 
contain high density regions. Thus the differences between the systems are not entirely
due to the differences in density.

\section{Discussion}
\subsection{Morphology of the superclusters}

We analyzed recently the morphology of individual superclusters, applying Minkowski 
functionals and shapefinders, and using the full density distribution in 
superclusters, at all density levels \citep{e07a,e08}. In these papers we used 
data from the 2dFGRS to delineate the richest superclusters and to study their 
morphology, as well as the morphology of individual galaxy populations.

Our studies included also the supercluster SCl~126, the richest supercluster in 
the SGW. Comparison of the results shows several differences and similarities. 
Due to sample limits a part of the supercluster SCl~126 was not covered by the 
2dFGRS. This is seen in Fig.~\ref{fig:scl6024sky} (left panel) that shows an 
extended filament at the declinations $\delta > 2^{\circ}$ not visible in the 
2dFGRS data. Due to this, the clumpiness of galaxies in this supercluster 
($V_3\leq 10$) in \citet{e08} was smaller than that found here; 
there are also small differences between  the morphological signatures of this 
supercluster found here and in   \citet{e08}. Both studies showed that the 
differences between the clumpiness of red and blue galaxies are larger in the 
outskirts of the supercluster and smaller in the core of the supercluster. This 
shows that the scale-dependent bias between the 2dFGRS and SDSS due to the $r$-band
selection of the SDSS \citep{cole} does not affect (strongly) the values of 
the Minkowski functionals. 

However, even in the present study, using the data from the SDSS DR7, we cannot be sure
that the SGW is fully covered by the observations. At low declinations, 
the SGW extends down to the DR7 limit, and it may be possible that the SGW continues
at lower declinations. Future observations are needed to see the total extent of the 
SGW.

Our present study showed that the peculiar morphology of the supercluster 
SCl~126 is especially strongly expressed in the core region of the supercluster. 
The morphology of another supercluster in the SGW, the supercluster SCl~111, 
resembles that of simulated superclusters studied in \citet{e07a}. This agrees 
with our comparison  of the morphology of individual richest superclusters 
\citep{e07a} in observations and in simulations that showed that the morphology 
of the simulated superclusters differs from the morphology of the supercluster 
supercluster SCl126. 

The morphology of the full supercluster SCl~126 can be described as a multibranching
filament while the core of this supercluster resembles a very rich filament
with only a small number of clumps at high density levels. This makes 
the core of the supercluster SCl~126 an extreme case of a cosmic filament.
The supercluster SCl~111, on the contrary, consists of several high density
clumps connected by lower density filaments. Such 
typical filaments between
rich clusters have lengths up to about 25\Mpc\ 
\citep{2010MNRAS.406.1609B,2010MNRAS.409..156B,2010MNRAS.406..320C}
although filaments with length of about 40\Mpc\ have also been
detected \citep{2004MNRAS.354L..61P}.
Even in the case of intracluster  filaments comparison with
model filaments shows discrepancies: while \citet{2010A&A...510A..38S}
show that model filaments tend to be shorter than observed filaments and they
do not form an extended network,
\citet{2010MNRAS.409..156B} find that filaments from
observations and models are in a good agreement. Thus even 
in the case of typical intracluster filaments future studies are
needed to compare further the properties of observed and model filaments.

Our analysis of substructures of superclusters as delineated by different galaxy 
populations revealed several differences and similarities between them. In the 
core of the supercluster SCl~126 (SCl~126c) the morphology of different galaxy 
populations is similar. In the outskirts of this supercluster the clumpiness of 
different galaxy populations is rather different. Here BRGs form some isolated 
clumps not seen in the clumpiness curves of other galaxy populations. The 
clumpiness of blue galaxies is small, blue galaxy population may have tunnels 
through them. 
Differences between the clumpiness of red and blue galaxies is the 
largest at intermediate mass fractions. In this respect our present results are 
similar to those obtained about this supercluster using the 2dFGRS data \citep{e08}. 
The differences between the clumpiness of red and blue galaxy populations in the 
supercluster SCl~111 are even larger than in the supercluster SCl~126.

Interestingly, in another rich supercluster, studied in 
\citet{e08}, the Sculptor supercluster, the clumpiness of blue galaxies was 
larger than the clumpiness of red galaxies. In \citet{e07a} we showed that the 
clumpiness of observed superclusters for bright and faint galaxies has a much 
larger scatter than that of simulated superclusters. This shows that 
superclusters which we have studied so far have a large diversity in how 
different galaxy populations determine their substructure. Explaining large 
variations in the distribution of different galaxy populations between 
superclusters is a challenge for galaxy formation models.

We calculated also the 
shape parameter $K_1/K_2$ (the ratio of shapefinders) for different galaxy 
populations in the superclusters SCl~126 and SCl~111, and showed 
that the shape parameters of different galaxy populations in the supercluster SCl~126
are similar. In the supercluster SCl~111 the shape parameter for red galaxies
has a smaller value than that for blue galaxies. A larger value of the shape parameter
indicates more planar systems. This calculation shows another difference between
these superclusters and suggests that the large-scale segregation
of red and blue galaxy populations in the supercluster SCl~111 is larger than
in the supercluster SCl~126. We note that  \citet{bas03} used the
shape parameter to characterize observed superclusters 
and to compare observed and simulated superclusters. 

A large clumpiness of the red subsystem and a small clumpiness of the blue subsystem 
shows that blue galaxies are distributed more homogeneously than red galaxies,
especially in the outskirts of superclusters. This is an evidence 
of a large scale morphological segregation of galaxies in superclusters,
discovered already in the first studies of superclusters and confirmed later
\citep[][among others]{1986ApJ...300...77G,2003MNRAS.339..652K,2005A&A...443..435W,
2006MNRAS.371...55H} 
and quantified using the correlation function
of galaxies at large scales 
\citep{1991MNRAS.252..261E,1991ApJ...382L...5G} 
(the ``2-halo'' contribution in halo models, see \citet{zehavi04,blake07}).

Earlier studies of the Minkowski functionals of the whole SDSS survey region
\citep[and references therein]{gott08} 
showed the SGW as a very strong density enhancement in the overall distribution
of galaxies, which causes a ``meatball'' shift of the genus of SDSS.
\citet{gott08} analyzed the topology of mock SDSS samples from the 
Millenium simulations, and 
concluded that such systems are not found in the simulations.  
Similar conclusions were reached by \citet{e06}.

This all shows that simulation of very rich superclusters is
complicated and present-day simulations do not yet explain all 
the features of the observed superclusters. Future N-body simulations 
for very large volumes and with more power at large scales are needed to 
model structures like the SGW more accurately than the present simulations
have done. 

The formation and evolution history of superclusters is a complex 
subject. The best way to follow their real evolution in time is to look 
for superclusters in deep surveys. In this respect an especially 
promising project is the ALHAMBRA Deep survey \citep{moles08} that 
will provide us with data about (possible) galaxy systems at very high 
redshifts, which can be searched for and analyzed using morphological 
methods. Comparing superclusters at different redshifts will clarify 
many questions about their evolution. A good template for such a comparison
is the recently confirmed distant ($z=0.55$) supercluster \citep{tanaka09}.

\subsection{BRGs and the first ranked galaxies}

The result that most BRGs lie in groups is consistent with the strong small-
scale clustering of the LRGs \citep{zehavi04,eisenstein04,blake07} that has been 
interpreted in the framework of the halo occupation model, where more massive 
haloes host a larger number of LRGs \citep{2008arXiv0805.0002V,zheng08,
2010ApJ...709...67T, 2010ApJ...709..115W}.
\citet{2008arXiv0805.0002V} showed also that more massive haloes host
more massive satellites, and satellites at smaller halocentric radius
and in more massive haloes are redder. \citet{zheng08} showed that,
according to the halo model, most LRGs are the first ranked galaxies
in groups, while in our sample about 60\% of BRGs are not the first ranked 
galaxies. This difference may be due to the use of nearby sample in our 
study.

The LRGs, selected so that they form almost a volume limited sample of galaxies 
up to the redshifts about $z = 0.5$, are mainly used as tracers of the large 
scale structure in the Universe at large distances. Our results show that in the 
supercluster SCl~126, rich groups with BRGs are located both in high density 
core and  in outskirts. In the supercluster SCl~111 rich groups with a larger 
number of BRGs can be found in high density regions, while in lower density 
regions there are poor groups with a small number of BRGs. Thus the brightest 
red galaxies trace the large scale structure in a somewhat different manner than 
other galaxies; these differences are larger in lower density regions of 
superclusters. 

We plan to study a 
larger sample of BRGs and superclusters to further understand the large scale 
distribution of BRGs, that
is important for
several purposes.
These galaxies (their more distant cousins, the LRGs) have been mainly used to study baryonic acoustic oscillations 
\citep{eisenstein2005, hutsi06, martinez09, 2010ApJ...710.1444K}. Many observational
projects to better measure these oscillations have been proposed and accepted. 
As an example, a new 
photometric galaxy redshift survey called PAU (Physics of the 
Accelerating Universe) will measure positions and redshifts for over 14 
million luminous red galaxies over 8000 square degrees in the sky, in 
the redshift range $0.1 < z < 0.9$ with a precision that is needed to 
measure baryon acoustic oscillations \citep{ben08}.

Recently, \citet{ho07} investigated the distribution of luminous red galaxies in 
47 distant X-ray clusters ($0.2 < z < 0.6$) and showed that the brightest LRGs 
are concentrated to the cluster centers, while a significant fraction of LRGs lies 
farther away from the centers. 
This agrees with our results about the large peculiar velocities
of BRGs.

We found that a large fraction of BRGs are spirals. Images of a few BRGs are 
shown in Fig.~\ref{fig:images}; images of all BRGs in the SGW are displayed in 
the web page \url{\texttt{http://www.aai.ee/$\sim$maret/SGWmorph.html}}, to show 
that our morphological classification of these bright galaxies is reliable. A 
recent analysis of colors and morphological types of galaxies in the Galaxy Zoo 
project \citep{bamford08, 2010MNRAS.405..783M} showed that, in fact, many spiral 
galaxies are red, moreover, massive galaxies are all red, independent of their 
morphology. Our results are in agreement with these much larger data. 
\citet{2009MNRAS.399..966S} showed that red spiral galaxies can be found in 
moderate dense environments and they are often satellite galaxies in the 
outskirts of haloes. Our results show, however,  that spiral red galaxies may be 
even the first ranked galaxies in groups. The scatter of colors of red spiral 
galaxies is large---a hint that their colors have changed recently in a group 
environment while their morphology has not changed yet?

In \citet{e08} we showed, using the data from the 2dFGRS, that in the supercluster
SCl~126 red galaxies with spectra typical of late-type galaxies are located
in poor groups at intermediate densities. In the present study we see that
red spiral galaxies reside also in rich groups in high density cores of superclusters,
being often the first ranked galaxies in groups. This means that the processes, which 
change the colors of spirals to red, must occur in both high
and low density regions. The same was concluded by \citet{2010MNRAS.405..783M}.
This, evidently, is a challenge for galaxy evolution scenarios \citep{kauff96}; 
we plan to explore
that, studying the detailed properties of individual SGW groups.

According to Cold Dark Matter models, cluster (group) haloes form hierarchically by 
merging of smaller mass haloes \citep{loeb2008}. If at present galaxy groups were 
virialized \citep[as shown, e.g. by][] {araya2009} and the galaxies in groups  
follow the gravitational potential, then we should expect  that the first ranked galaxies of 
groups would lie at the centers of groups and have small peculiar velocities (regardless 
of the group richness or its velocity dispersion) \citep{ost75,merritt84,malumuth92}. 
However, our results show that a significant fraction of 
first ranked  galaxies in groups have large 
peculiar velocities. 
Our results are consistent with those by \citet{coziol09} 
based on the data about rich clusters of galaxies but
do not fit well into the picture where the brightest galaxies
in groups are considered as the central galaxies 
\citep{2005MNRAS.356.1293Y}.
The large peculiar velocities of the first ranked
galaxies in groups indicate that all galaxy groups are not virialized 
yet. \citet{tov09} arrived at the same conclusion on the basis of the analysis of poor 
groups, using, in particular, our group catalogue based 
on the SDSS DR5 \citep{tago08}. \citet{niemi} showed that a 
significant fraction of nearby groups of galaxies are not gravitationally bound 
systems. This is important because  the masses of observed groups 
are often estimated assuming that the groups are bound.

We showed that in the core of the richest supercluster in the SGW, the 
supercluster SCl~126,  the peculiar velocities of elliptical BRGs 
are large. 
One possible reason for that may be that (rich) 
groups there have formed recently by merging 
of poorer groups consisting mostly of red elliptical galaxies. 
In Fig.~\ref{fig:images}, the left panel shows the image of the first ranked galaxy of the 
richest group in the core of the supercluster SCl~126 that demonstrates clear signs of 
merging (the two brightest galaxies in this group have a common red halo). This group 
is the Abell cluster A1750, where available optical and X-ray data 
show evidence of merging \citep{don,bel04,hwang09}, supporting our assumption. 
In addition, \citet{2010ApJ...709...67T} in their analysis of the distribution of distant 
red galaxies in the halo model framework mention that distant red galaxies must have 
formed their stars before they become satellites in haloes. This is possible if 
these galaxies were members of poorer groups before the groups merged into a richer one. 
The scatter of colors of elliptical BRGs in the groups in the core of the 
supercluster SCl~126 is small. 
This suggest that even if the rich groups in the SCl~126c formed recently, their galaxies 
obtained their colors when residing in poor groups, before the merger of the
groups.

In the core of the supercluster SCL~126 the scatter of colors of spiral BRGs 
is larger and the peculiar velocities of them are smaller than those of 
elliptical BRGs. A detailed study of groups in different superclusters
of the SGW is needed to understand these differences.  

In the outskirst of the supercluster SCl~126 and in the supercluster
SCl~111, spiral BRGs have larger peculiar velocities than 
elliptical BRGs. 

Evolution of galaxy groups depends on the environment, being faster in the high 
density cores of superclusters \citep[see, e.g.][]{tempel09}. This may partly 
explain the differences between the properties of different superclusters in the 
SGW, where the highest global densities are found in the supercluster 
SCl~126.

One unexpected result of our study is that in the richest 
superclusters in the SGW, SCl~126 and SCl~111, groups with an elliptical first ranked 
BRGs are located more uniformly than groups with a spiral first ranked BRG.  
Groups with an elliptical BRG as the first ranked galaxy 
are richer, and rich groups should be concentrated in the denser parts of the 
supercluster. We plan to study this difference in more detail, 
using a larger sample of superclusters. 

Of course, our results depend on how good is our group 
catalogue. As we mentioned, in order to take into 
account selection effects, we carefully rescaled the linking parameter with 
distance. As a result, the maximum projected sizes in the sky and velocity 
dispersions of our groups in our catalogue are similar at all distances 
\citep[T10]{tago08}. In the group catalogue the group richnesses decrease 
rapidly starting from $D=300$~\Mpc, due to the use of a flux-limited galaxy sample. 
We used the same distance limit,  to minimize 
selection effects, which still might affect our results. A good agreement with other 
results \citep[e.g][]{coziol09} suggest that our choice of parameters to 
select groups has been reliable.

\section{Conclusions}

We used the fourth Minkowski functional $V_3$ and the morphological signature 
$K_1$-$K_2$ to characterize the morphology of the SGW, the richest large-scale
galaxy system in the nearby Universe, at a series of density levels and
to study the morphology of the galaxy populations in the SGW. 
We analyzed the group membership of BRGs, compared the properties of 
groups with and without BRGs in different superclusters 
of the SGW, and studied
the colors, luminosities, morphological types and peculiar
velocities of BRGs and those first ranked galaxies, which do not
belong to BRGs.
Our main conclusions are as follows.

\begin{itemize}
\item
We demonstrated that the morphology of the SGW varies with the overall
density level. Starting from the highest densities, it changes from a morphology 
characteristic of a simple filament to that of a multibranching filament 
and then becomes 
characteristic of a very clumpy, planar system. 
The changes in morphology suggest that the right density level to extract 
individual superclusters from the overall (luminosity) density field
is $D = 4.9$. A lower density level, 
$D = 4.7$, can be used to select the whole SGW. 

At all density levels the morphological signature changes at a mass fraction 
$m_f \approx 0.7$---a change of morphology at a crossover from the outskirts to 
the core regions of superclusters. 
\item In the core of the supercluster SCl~126 (the richest supercluster in the SGW)
the clumpiness of red and blue galaxies, as well as BRGs, is similar, 
with the maximum value of the 4th Minkowski functional $V_3 = 9$.

In the outskirts of the supercluster SCl~126, as well as in the supercluster
SCl~111 (another rich supercluster in the SGW),  
the clumpiness of red galaxies is larger than
the clumpiness of blue galaxies (in other words, the distribution of blue galaxies
is more homogeneous than that of red galaxies). The clumpiness of BRGs is similar to that
of red galaxies. The systems of blue galaxies may have tunnels through them
at intermediate mass fractions. 
\item
The morphology of the supercluster SCl~126 and especially it's core region
differs from the morphology of the very rich simulated superclusters 
studied in \citet{e07a}.
The morphology of the supercluster SCl~111 
is similar to that of poorer simulated superclusters. This shows that
 the peculiar morphology of the SGW comes from the unusual morphology
 of the supercluster SCl~126.

\item  We assessed the statistical significance of the differences between
the 4th Minkowski functional $V_3$ and morphological signatures
of different galaxy populations in superclusters using 
halo model and smoothed bootstrap
and showed that at least at some mass fraction intervals (depending
on the populations) the differences between galaxy populations
are statistically significant at
least at the 95\% confidence level.

\item There are large variations between superclusters
 of how different galaxy 
populations determine their substructure. Explaining this diversity in the 
distribution of different galaxy populations between superclusters is a 
challenge for galaxy formation models. 

\item A significant fraction of BRGs lies in groups; there are groups 
with at least 5 BRGs in them. About 40\% of BRGs are the first ranked 
galaxies in groups. Groups with BRGs are richer, more luminous and
have larger velocity dispersions than groups without BRGs. In the supercluster
SCl~126 groups with a large number of BRGs are located both  in the
core and in outskirts of the supercluster, while in the supercluster
SCl~111 groups with a large number of BRGs lie in high density regions only.
The fraction of isolated BRGs in the core 
of the supercluster SCl~126 is about half of that in the outskirts of this
supercluster and in the supercluster SCl~111.
 
\item About half of the BRGs and those first ranked galaxies which do not
belong to BRGs have 
large peculiar velocities in respect to the group's center.
The peculiar velocities of BRGs and the first ranked galaxies 
are larger in the supercluster SCl~126 than those
in the supercluster SCl~111, showing that the groups  
in the supercluster SCl~126 are dynamically more active 
(actively merging) than those in the supercluster SCl~111.
\item
In the core of the supercluster
SCl~126 (SCl~126c) the fraction of red galaxies among those first
ranked galaxies which are not BRGs is larger 
and the fraction of spirals 
among them is smaller than in the outskirts
of this supercluster and in the supercluster SCl~111.
\item About 1/3 of the BRGs in the two richest supercluster in the SGW
(SCl~126 and SCl~111) are spirals. In the color-magnitude diagram the 
scatter of colors of red elliptical BRGs is smaller than that of red spiral 
BRGs. There are some blue galaxies among the BRGs. Most of blue BRGs  
are spirals. The peculiar velocities of elliptical BRGs in 
the core region of the supercluster SCl~126 are larger than
the peculiar velocities of spiral BRGs, while in other superclusters in 
the SGW spiral BRGs have larger peculiar velocities than elliptical
BRGs. The spatial distribution of groups with elliptical or spiral BRG as the first ranked 
galaxy in superclusters is also different.

\end{itemize}

The fourth Minkowski functional and the 
morphological signature of the galaxy populations, as well as the properties
of groups and galaxies  in 
different superclusters in the SGW differ from each other, demonstrating
that the formation and evolution of different superclusters in the SGW
have been different. 

While there are several superclusters as the SCl~111 \citep{e07a}, and 
model superclusters resemble these \citep[see, e.g.][]{ara08}, 
the supercluster SCl~126 is an exceptional supercluster in many aspects.
While the SCl~111 is more relaxed and its dynamical evolution has been
almost finished (as it should be in the case of accelerated expansion),
the SCl~126 is yet dynamically active. The initial conditions for its volume
must have carried considerably higher density levels and corresponding velocity perturbations
than in the neighbouring regions (e.g., SCl~111). This is seen both in the
overall large-scale density distribution, in its morphology, and in the
group and galaxy content. This might be an argument for non-Gaussian initial
perturbations, as suggested by \citet{1996ApJ...462L...1G}.
This all shows that the full SGW 
is an arrangement
of several superclusters with different evolutional history.
New simulations in larger volumes are needed
to study the morphology of superclusters and the evolution of the morphology
of simulated superclusters to understand whether the exceptional morphology
of the supercluster SCl~126 can be explained by simulations.

We continue our study of the properties of groups in the SGW in more detail in 
another paper, where we  also analyze the richest groups in the SGW, their 
substructure and galaxy populations \citep{2010A&A...522A..92E}. We plan to 
continue the study of more distant BRGs in order to understand how strongly our 
present results about BRGs are affected by possible selection effects. The present 
study was focused on the properties of the SGW and galaxy populations in the 
SGW. Several our results show a need to continue the study of the morphology of 
a large sample of 
superclusters (Einasto et al. 2011, A\&A, accepted), 
especially the evolution of the morphology of the large-scale structure
in simulations and also in observations, using deep surveys.
We also plan continue to study the properties of groups and galaxies in superclusters
and in the voids between them, using larger and deeper samples. 
\section*{Acknowledgments}

We thank the anonymous referee for a very detailed reading of the
manuscript and for many useful comments that helped us to improve the
original manuscript. 
We thank Michael Blanton for explaining us how to deal
with fiber collisions in the SDSS.
The present study was supported by the Estonian Science Foundation
grants No. 8005, 7146 and 7765, and by the Estonian Ministry for Education and
Science grant SF0060067s08. It has also been supported by the University of Valencia 
(Vicerrectorado de Investigaci\'on) through a
visiting professorship for Enn Saar and by the Spanish MEC projects
``ALHAMBRA'' (AYA2006-14056), ``PAU'' (CSD2007-00060), including
FEDER contributions, and the Generalitat Valenciana project of excellence 
PROMETEO/2009/064. J.E.  thanks
Astrophysikalisches Institut Potsdam (using DFG-grant 436 EST 17/4/06), 
where part of this study was performed.  
The density maps and the supercluster catalogues were
calculated at the High Performance Computing Centre, University of Tartu.

We thank the SDSS Team for the publicly available data
releases.  
Funding for the SDSS and SDSS-II has been
provided by the Alfred P. Sloan Foundation, the Participating
Institutions, the National Science Foundation, the U.S. Department of
Energy, the National Aeronautics and Space Administration, the
Japanese Monbukagakusho, the Max Planck Society, and the Higher
Education Funding Council for England. The SDSS Web Site is
\texttt{http://www.sdss.org/}.

The SDSS is managed by the Astrophysical Research Consortium for the
Participating Institutions. The Participating Institutions are the
American Museum of Natural History, Astrophysical Institute Potsdam,
University of Basel, University of Cambridge, Case Western Reserve
University, University of Chicago, Drexel University, Fermilab, the
Institute for Advanced Study, the Japan Participation Group, Johns
Hopkins University, the Joint Institute for Nuclear Astrophysics, the
Kavli Institute for Particle Astrophysics and Cosmology, the Korean
Scientist Group, the Chinese Academy of Sciences (LAMOST), Los Alamos
National Laboratory, the Max-Planck-Institute for Astronomy (MPIA),
the Max-Planck-Institute for Astrophysics (MPA), New Mexico State
University, Ohio State University, University of Pittsburgh,
University of Portsmouth, Princeton University, the United States
Naval Observatory, and the University of Washington.

\bibliographystyle{apj}
\bibliography{sgwe7morf.bib}

\begin{appendix}
\section{Density field and superclusters}
\label{sec:DF}

\subsection{Kernel method}

To calculate a luminosity density field, 
we must convert the spatial positions of galaxies $\mathbf{r}_i$ 
and their luminosities  $L_i$ into
spatial (luminosity) densities. The standard approach is to use kernel densities
\citep{silverman86}:
\begin{equation}
    \rho(\mathbf{r}) = \sum_i K\left( \mathbf{r} - \mathbf{r}_i; a\right) L_i,
\end{equation}
where the sum is over all galaxies, and $K\left(\mathbf{r};
a\right)$ is a kernel function of a width $a$. Good kernels
for calculating densities on a spatial grid are generated by box splines
$B_J$. Box splines are local and they are interpolating on a grid:
\begin{equation}
    \sum_i B_J \left(x-i \right) = 1,
\end{equation}
for any $x$ and a small number of indices that give non-zero values for $B_J(x)$.
We use the popular $B_3$ spline function:
\begin{equation}
\label{eq:b3}
    B_3(x) = \frac{1}{12} \left(|x-2|^3 - 4|x-1|^3 + 6|x|^3 - 4|x+1|^3 + |x+2|^3\right).
\end{equation}
We define the (one-dimensional) $B_3$ box spline kernel $K_B^{(1)}$ of the width $a$ as
\begin{equation}
    K_B^{(1)}(x;a,\delta) = B_3(x/a)(\delta / a),
\end{equation}
where $\delta$ is the grid step. This kernel differs from zero only
in the interval $x\in[-2a,2a]$; it is close to a Gaussian with $\sigma=1$ in the
region $x\in[-a,a]$, so its effective width is $2a$ \citep[see, e.g.,][]{saar09}.
The kernel preserves the
interpolation property exactly for all values of $a$ and $\delta$,
where the ratio $a/\delta$ is an integer. (This kernel can be used also if this ratio
is not an integer, 
and $a \gg \delta$; the kernel sums to 1 in this case, too, with a very small error).
This means that if we apply this kernel to $N$ points on a one-dimensional grid,
the sum of the densities over the grid is exactly $N$.
 
The three-dimensional kernel $K_B^{(3)}$
is given by the direct product of three one-dimensional kernels:
\begin{equation}
    K_B^{(3)}(\mathbf{r};a,\delta) \equiv K_3^{(1)}(x;a,\delta) K_3^{(1)}(y;a,\delta) K_3^{(1)}(z;a,\delta),
\end{equation}
where $\mathbf{r} \equiv \{x,y,z\}$. Although this is a direct product,
it is isotropic to a good degree \citep{saar09}.

\subsection{Density fields}
To use galaxy data, we applied first the $k+e$-correction
to each galaxy (Sect.~\ref{subsec:pops}). Also, we have to consider
the luminosities of the galaxies that lie outside the observational
window of the survey. Assuming that every galaxy is a visible member of a
density enhancement (a group or cluster), we estimate
the amount of unobserved luminosity and weigh each galaxy accordingly:
\begin{equation}
    L_{w} = W_L(d)\; L_{obs}.
\end{equation}
Here $W_L(d)$ is a distance-dependent weight:
\begin{equation}
    W_L(d) = \frac{\int_0^\infty L\;F(L)dL}{\int_{L_1(d)}^{L_2(d)} L\;F(L) dL},
\end{equation}
where $F(L)$ is the luminosity function and $L_1(d)$ and $L_2(d)$ are the
luminosity window limits at a distance $d$.

\begin{figure}[ht]
\centering
\resizebox{0.40\textwidth}{!}{\includegraphics*{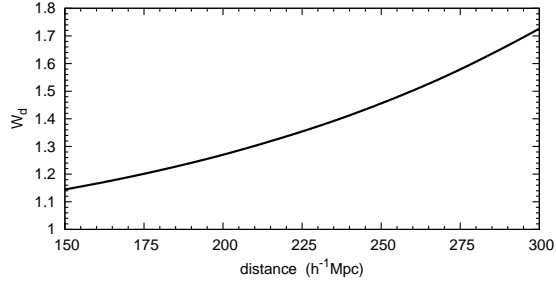}}
\caption{The weights used to correct for probable group members outside the
  observational luminosity window.}
\label{fig:weight}
\end{figure}

The luminosity weights for the groups of the SDSS DR7 in the region of the 
SGW are plotted as a
function of the distance from the observer in Fig.~\ref{fig:weight}.  We see
that the mean weight is slightly higher than unity (about 1.4)
within the sample limits, $150\ge D_{com}\le 300$~\Mpc.  At small distances very bright member galaxies of
groups are missing, they lie outside the observational window.
When the distance is larger, the weights
increase due to the absence of faint galaxies.  

Using the rms velocity $\sigma_v$, translated into distance,
and the rms projected radius $\sigma_r$ from the group catalogue
(T10),
we supress the cluster finger redshift distortions. We
divide the radial distances between the group galaxies and the group centre by the ratio
of the rms sizes of the group finger:
\begin{equation}
    d_{gal,f} = d_{group} + (d_{gal,i} - d_{group})\; \sigma_r / \sigma_v.
\end{equation}
This removes the smudging effect the fingers have on
the density field.

The densities were calculated on a cartesian grid based on the SDSS $\eta$,
$\lambda$ coordinate system, as it allowed the most efficient fit of
the galaxy sample cone into a brick.
The grid coordinates were found as follows:
\begin{eqnarray}
\label{eq.xyz}
x&=&-d_{gal} \sin\lambda,\nonumber\\
y&=&d_{gal} \cos\lambda \cos\eta,\\
z&=&d_{gal} \cos\lambda \sin\eta.\nonumber
\end{eqnarray}
We used an 1 Mpc/$h$ step grid and chose the kernel width $a=8$ Mpc/$h$.
As we noted above, this means that the
kernel differs from zero within the radius 16 Mpc/$h$,
but significantly so only inside the 8 Mpc/$h$ radius.

Before extracting superclusters we apply the DR7 mask 
constructed by P.~Arnalte-Mur
\citep{martinez09} to the density field
and convert densities into units of mean density. The mean
density is defined as the average over all pixel values inside the mask. The mask is
designed to follow the edges of the survey and the galaxy distribution inside the
mask is assumed to be homogeneous.

\subsection{Superclusters}

We create a set of density contours by choosing a relative density threshold $D$ 
and define 
connected volumes above this threshold
as superclusters. We assemble galaxy superclusters by collecting all the
galaxies that are located in the supercluster volume. 

Different thresholds yield different supercluster catalogues. A supercluster 
catalogue combines both the density field and galaxy information. We find the 
supercluster location, volume, diameter, total luminosity, the numbers of galaxies and 
groups. For a supercluster ID we use the coordinates of  a ``marker galaxy'' 
that we choose to be a bright galaxy nearest to the highest density peak in the 
supercluster volume (often this is the first ranked galaxy in the most luminous 
group of a supercluster).

\section{Minkowski functionals and shapefinders} 
\label{sec:MF}

Consider an
excursion set $F_{\phi_0}$ of a field $\phi(\mathbf{x})$ (the set
of all points where the density is larger than a given limit,
$\phi(\mathbf{x}\ge\phi_0$)). Then, the first
Minkowski functional (the volume functional) is the volume of 
this region (the excursion set):
\begin{equation}
\label{mf0}
V_0(\phi_0)=\int_{F_{\phi_0}}d^3x\;.
\end{equation}
The second Minkowski functional is proportional to the surface area
of the boundary $\delta F_\phi$ of the excursion set:
\begin{equation}
\label{mf1}
V_1(\phi_0)=\frac16\int_{\delta F_{\phi_0}}dS(\mathbf{x})\;,
\end{equation}
(but it is not the area itself, notice the constant).
The third Minkowski functional is proportional to the
integrated mean curvature 
$C$ of the boundary:
\begin{equation}
\label{mf2}
V_2(\phi_0)=\frac1{6\pi}\int_{\delta F_{\phi_0}}
    \left(\frac1{R_1(\mathbf{x})}+\frac1{R_2(\mathbf{x})}\right)dS(\mathbf{x})\;,
\end{equation}
where $R_1(\mathbf{x})$ and $R_2(\mathbf{x})$ 
are the principal radii of curvature of the boundary.
The fourth Minkowski functional is proportional to the integrated
Gaussian curvature (the Euler characteristic) 
of the boundary:
\begin{equation}
\label{mf3}
V_3(\phi_0)=\frac1{4\pi}\int_{\delta F_{\phi_0}}
    \frac1{R_1(\mathbf{x})R_2(\mathbf{x})}dS(\mathbf{x})\;.
\end{equation}
At high (low) densities this functional gives us the number of isolated 
clumps (void bubbles) in the sample 
\citep{mar03,saar06}:

\begin{equation}
\label{v3}
V_3=N_{\mbox{clumps}} + N_{\mbox{bubbles}} - N_{\mbox{tunnels}}.
\end{equation}.

As the argument labeling the isodensity surfaces, we chose the (excluded) mass
fraction $m_f$---the ratio of the mass in the regions with the density {\em lower}
than the density at the surface, to the total mass of the supercluster. When
this ratio runs from 0 to 1, the iso-surfaces move from the outer limiting
boundary into the center of the supercluster, i.e. the fraction $m_f=0$
corresponds to the whole supercluster, and $m_f=1$---to its highest density
peak.

We use directly only the fourth Minkowski functional in this paper;
the other functionals are used to calculate the shapefinders
\citep{sah98,sss04,saar09}. 
The shapefinders are defined as a
set of combinations of Minkowski functionals: $H_1=3V/S$ (thickness),
$H_2=S/C$ (width), and $H_3=C/4\pi$ (length).  The
shapefinders have dimensions of length and are normalized to give $H_i=R$
for a sphere of radius $R$.  For a 
smooth (ellipsoidal) surfaces, the shapefinders $H_i$
follow the inequalities $H_1\leq H_2\leq H_3$.  Oblate ellipsoids (pancakes)
are characterized by $H_1 << H_2 \approx H_3$, while prolate ellipsoids
(filaments) are described by $H_1 \approx H_2 << H_3$.

\citet{sah98} also defined  two dimensionless
shapefinders $K_1$ (planarity) and $K_2$ (filamentarity): 
$K_1 = (H_2 - H_1)/(H_2 + H_1)$ and $K_2 = (H_3 -
H_2)/(H_3 + H_2)$.

If systems under study consists of 
a number of small clumps (as the supercluster cores at high
mass fractions) then the surfaces have complex shape and the 
inequalities $H_1\leq H_2\leq H_3$ may not be valid. In that case it is possible
that the values of shapefinders $K_1$ and $K_2$ become negative.
This is seen, for example, 
in Fig~\ref{fig:MFG1} - Fig~\ref{fig:MFG3} and 
in Fig.~14 of  \citet{e08} where we modelled the
substructure  of superclusters using empirical models where
superclusters contained a large number of small groups. 

In the $(K_1,K_2)$-plane filaments are located near the $K_2$-axis,
pancakes near the $K_1$-axis, and ribbons along the diagonal, connecting 
the spheres at the origin with the ideal ribbon at $(1,1)$. 

\section{Morphological templates}
\label{sec:MT}

In previous papers \citep{e07a,e08} we generated a series of empirical models, which 
served us as morphological templates to understand the clumpiness of the 
superclusters and the behaviour of the shapefinders. In these models, we did not 
study the possibility that superclusters have tunnels through them, as suggested by the 
negative values of the 4th Minkowski functional $V_3$ in Fig.~\ref{fig:MFG2} and 
Fig.~\ref{fig:MFG3}. For this appendix, we generated an additional series of empirical 
models to understand better the substructure of superclusters, in particular, 
the effect of tunnels. These tunnels 
may appear at intermediate density levels where some galaxies do not 
contribute any more to a supercluster. 

\begin{figure*}[ht]
\centering
\resizebox{0.45\textwidth}{!}{\includegraphics*{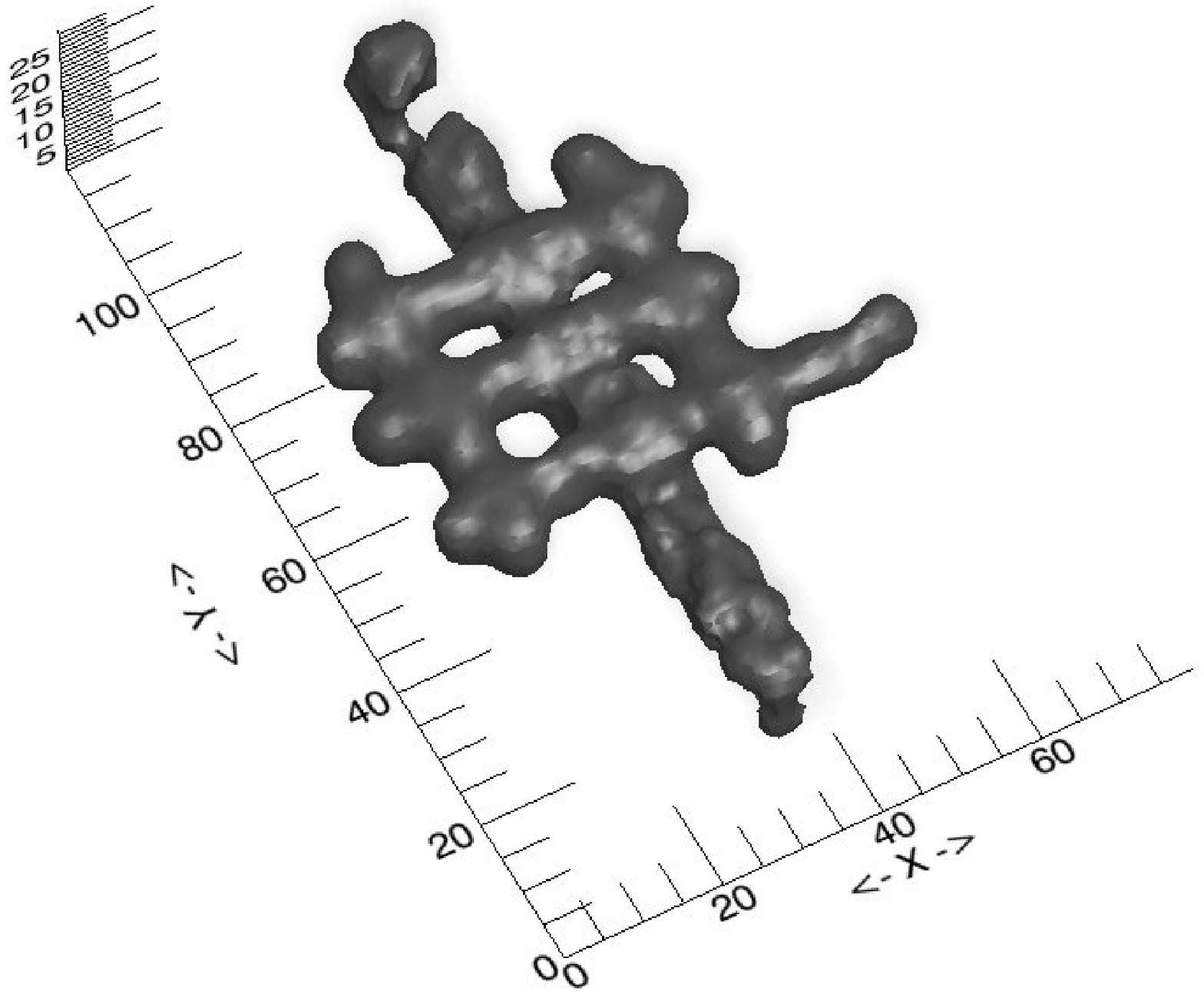}}
\resizebox{0.45\textwidth}{!}{\includegraphics*{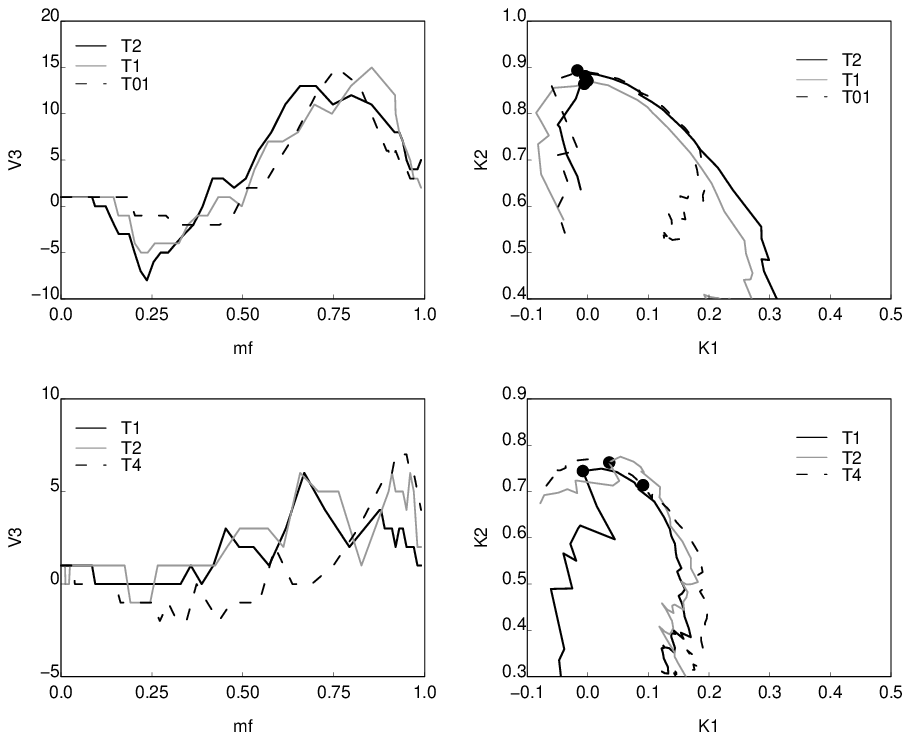}}
\hspace*{2mm}\\
\caption{
Morphological templates: crossing filaments forming tunnels (a ``forked pattern''
in knitting). Left panel: the most complex supercluster template T4.
Middle panels: the Minkowski functional $V_3$, the right panels---the
morphological signature for the templates described in the text. 
The template number (as in T1) shows the number of tunnels in the model 
(1, 2 or 4, while 01 denotes a model
with one wide tunnel). In all models,
one filament is long and thick (the main filament or the body of the supercluster).
In the upper panels, other filaments are also long but thin, in the lower panels 
one more filament is also long (but thin); this filament is crossing 
the main filament. Other filaments are short (both crossing and parallel).
}
\label{fig:tunnel}
\end{figure*}

In these models, the overall distribution 
of points (that mimic individual galaxies) resembles a filament 
with a size $10\times 20\times 100$ (in grid units). We add 
additional filaments into the model, some of them are parallel to the
main filament, others are crossing to form  holes (tunnels)
between filaments. (This configuration of filaments 
looks like a pattern known as a ``forked pattern'' in knitting). 
The total number of points in a main filament is always 500,
in additional filaments---200 
(approximately the number of galaxies in our observed superclusters,
and in their individual galaxy populations). 
The size of additional filaments is  $5\times 5\times 100$ (in grid units)
(long filaments, the first row in Fig.~\ref{fig:tunnel}). In the lower 
panels, 
the length  of the short filaments is $40$ and the length of
a long filament---$60$ (in grid units).
Where filaments overlap, high density regions form that
mimic density enhancements (clusters) inside superclusters.
We plot the Euler characteristics and morphological signatures for
these models in Figs.~\ref{fig:tunnel}. 

Fig.~\ref{fig:tunnel} shows that in models with tunnels, the value of 
the 4th Minkowski functional $V_3$ becomes negative. The exact value of 
$V_3$ depends also on the number of clumps in the model. The morphological 
signature $K_1$-$K_2$ depends on the length of the filaments. 
Due to too  long additional filaments the planarities $K_1$
in the upper right panel of Fig.~\ref{fig:tunnel} are too large.
When additional filaments are short, then the planarities $K_1$
are smaller and the morphological signature $K_1$-$K_2$ of the model
is rather similar to that of real superclusters. Such models mimic the
superclusters where the rich systems (groups and clusters of galaxies)
are connected by fainter systems leaving tunnels between them 
(as in the supercluster SCl~111) or forming a
multibranching filament (as the supercluster SCl~126).

\section{Estimating local density distributions for a Cox process}
\label{sec:Cox}

\begin{figure*}[ht]
\centering
\resizebox{0.54\textwidth}{!}{\includegraphics*{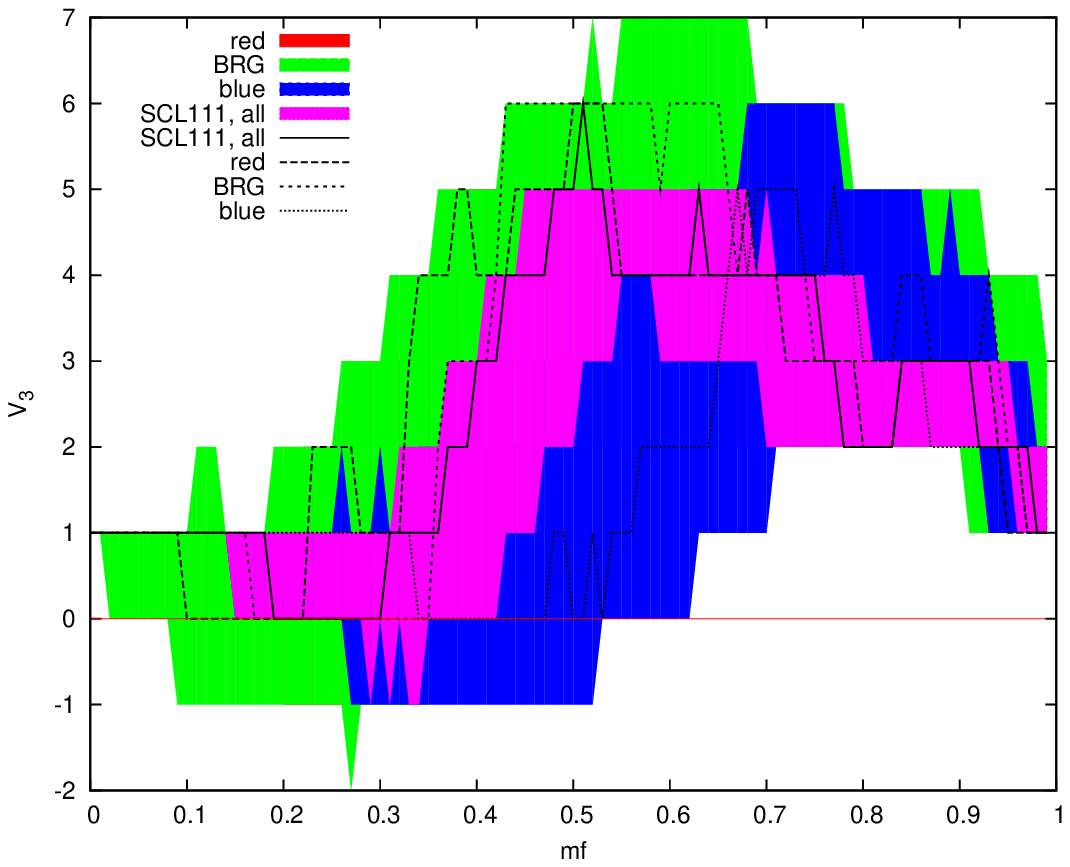}}
\resizebox{0.262\textwidth}{!}{\includegraphics*{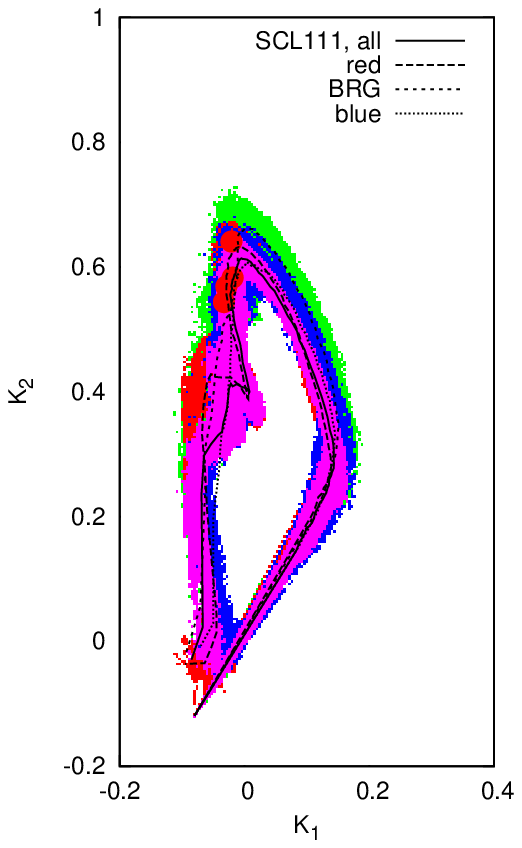}}
\\
\caption{
The 4th Minkowski functional $V_3$ (left panel) and the shapefinders 
 $K_1$ (planarity) and $K_2$ (filamentarity) 
 (right panel) for the red and blue galaxies and the BRGs in  
the supercluster SCl~111. 
The black line denotes  $V_3$ for all galaxies, dashed line with long
dashes -- for red galaxies, 
dashed line with short dashes -- for BRG-s, and dotted line for blue galaxies.
The filled circles in the right panel mark the value of the mass fraction 
$m_f\approx  0.7$. The colored regions
show the 95\% confidence regions
obtained with the shrunk smoothed bootstrap. With red color
we show the confidence regions for red galaxies, blue corresponds
to blue galaxies, green to BRGs and magenta to the full supercluster.
}
\label{fig:app-scl111}
\end{figure*}

A popular model for the galaxy distribution is Cox point process  \citep[see, 
e.g.][]{mar03}, where galaxies form a Poisson point process with spatially 
varying intensity, which is given by a realization of a random field. This model 
has been used successfully to estimate the moments of the galaxy distribution 
(correlation functions and spectral densities) and to explore their sampling 
properties. As the spatial density distribution in a finite region is similar to 
probability distribution, we can use the methods developed to estimate pointwise 
errors of probability distributions. In case we do not have a model for the 
distribution, the popular way is to use bootstrap. As is well known, the 
standard bootstrap -- selecting supercluster galaxies randomly with replacement 
-- is not suitable for estimating probability densities, but the smoothed 
bootstrap is \citep{silverman87,davison2009}. For kernel densities, as used in 
this paper, the smoothed bootstrap starts as the standard bootstrap by selecting 
a galaxy (with replacement), but then adding a random shift with the same 
distribution as the kernel to its coordinates. In  order to retain the 
covariance structure of the density, the shifts are reduced by 
$(1+h^2/\sigma^2)^{1/2}$, where $h$ is the original kernel width and $\sigma$ is 
the rms coordinate error for the data \citep{silverman87}. This is called shrunk 
smoothed bootstrap.

We used such bootstrapped galaxy sample realizations to estimate the confidence 
regions for the third Minkowski functional $V_3$ and for the morphological 
signature in the $K_1$-$K_2$ plane. As an example, we show the results in 
Fig.~\ref{fig:app-scl111} for the supercluster SCl~111 (1000 realizations; the 
results for the other two superclusters studied in this paper are similar). We 
see that the 95\% confidence regions are very wide and those for different 
populations practically overlap, both for $V_3$ and for the morphological 
signature. Does it mean that there is really no inner structure in this 
supercluster, and no differences in the spatial distribution of different galaxy 
populations? We see here the evidence that the Cox process model is not good for 
describing the detailed structure of galaxy associations. It does not account 
for the scale dependence of the galaxy distribution properties, and a Poisson 
process clearly cannot describe in full the complex process of galaxy formation 
that leaves its traces in the final spatial distribution of galaxies. The halo-
model based approach used in the main text reflects better the actual observed 
galaxy distribution.

\end{appendix}

\end{document}